\documentclass[fleqn,usenatbib]{mnras}

\usepackage{newtxtext,newtxmath}
\usepackage{graphicx}
\usepackage{amsmath} 
\usepackage{gensymb}
\usepackage{tabularx}
\usepackage{mathtools}
\usepackage{hyperref}
\usepackage{placeins}
\usepackage{lineno}

\usepackage[T1]{fontenc}
\DeclareRobustCommand{\VAN}[3]{#2}
\let\VANthebibliography\thebibliography
\def\thebibliography{\DeclareRobustCommand{\VAN}[3]{##3}\VANthebibliography}

\title[Amoeba]{Amoeba: An AGN Model of Optical Emissions Beyond steady-state Accretion discs}

\author[H. Best]{
Henry J. Best,$^{1, 2, 3, 4}$\thanks{E-mail: hbest@gradcenter.cuny.edu}
Matthew O'Dowd,$^{1, 2, 3}$
Joshua Fagin,$^{1, 2, 3}$ 
James H. H. Chan,$^{1, 2, 3}$ 
and Bridget Ierace$^{1, 2, 3}$ \\
$^{1}$The Graduate Center of the City University of New York, 365 Fifth Avenue, New York, NY 10016, USA\\
$^{2}$Department of Astrophysics, American Museum of Natural History, Central Park West and 79th Street, NY 10024-5192, USA\\
$^{3}$Department of Physics and Astronomy, Lehman College of the CUNY, Bronx, NY 10468, USA\\
$^{4}$Department of Theoretical Physics and Astrophysics, Faculty of Science,
Masaryk University, Kotlářská 2, CZ-611 37 Brno, Czech Republic
}

\date{Accepted XXX. Received YYY; in original form ZZZ}

\pubyear{20xx}

\begin{document}
\label{firstpage}
\pagerange{\pageref{firstpage}--\pageref{lastpage}}
\maketitle


\begin{abstract}
\label{Abstract}
Active Galactic Nuclei (AGN) are objects located in the heart of galaxies which emit powerful and complex radiation across the electromagnetic spectrum.
Understanding AGN has become a topic of interest due to their importance in galactic evolution and their ability to act as a probe of the distant Universe.
Within the next few years, wide-field surveys such as the Legacy Survey of Space and Time (LSST) at the Rubin Vera Observatory are expected to increase the number of known AGN to $\mathcal{O} (10^{7})$ and the number of strongly lensed AGN to $\mathcal{O} (10^{4})$.
In this paper we introduce \href{https://github.com/Henry-Best-01/Amoeba}{\texttt{Amoeba}}: an AGN Model of Optical Emission Beyond steady-state Accretion discs.
The goal of \texttt{Amoeba} is to provide a modular and flexible modelling environment for AGN, in which all components can interact with each other.
Through this work we describe the framework for major AGN components to vary self-consistently and keep flux distributions to connect these components to spatial dependent processes.
We model properties beyond traditional single-component models, such as the reverberation of the corona's bending power law power spectrum through the accretion disc, modulated by the diffuse continuum, and then propagated through the broad line region (BLR).
We simulate obscuration by the dusty torus and differential magnification of the disc and BLR due to microlensing.
These features are joined together to create some of the most realistic light curve simulations to date.
\texttt{Amoeba} takes a step forward in AGN modelling by joining the accretion disc, diffuse continuum, BLR, torus, and microlensing into a coherent macro-model.
\end{abstract}


\maketitle

\section{Introduction}
\label{IntroductionSection}

Active Galactic Nuclei (AGN) are believed to be supermassive black holes (SMBHs) actively accreting material from the host galaxy~\citep{Salpeter64, Zeldovich64}.
This process releases gravitational energy as intense radiation across the electromagnetic spectrum~\citep[See][for a review]{Padovani17}.
Quasars are a particular subclass of AGN which have a bright continuum, likely resulting from an unobscured accretion disc.
These can outshine their host galaxy and have been observed at z > 7~\citep{Mortlock11, Banados18}.
AGN are invaluable tools for probing cosmological phenomena due to their luminous and stochastic emission.
This variability allows us to probe AGN structure, AGN-galaxy co-evolution, and the Universe at cosmological scales~\citep[e.g.][]{Urry95, Fabian12, Birrer22}.

Recently, gravitationally lensed AGN have become a topic of interest to study both the AGN structure and cosmology.
Gravitational lensing is a prediction of general relativity (GR), where the gravitational field of a massive object bends the trajectory of light.
This may magnify, distort, and even create multiple images of the background source~\citep[see][for reviews]{Wambsganss, Bartelmann2010}.
The natural magnification from this lensing can allow for the study of distant sources which would be otherwise fall below limiting magnitudes, while the distortions and multiplicity of images can inform us of the distribution of matter in the Universe~\citep[e.g.][]{Courbin02}.
The most extreme case of gravitational lensing which creates multiple images is known as strong lensing. 

Strongly lensed AGN are particularly useful for their ability to place constraints on the Hubble constant $H_{0}$ through \emph{time delay cosmography}~\citep[see][for a review]{Birrer22}.
Time delay cosmography relies on two components, lens modelling and time delay measurements~\citep{Refsdal64b}.
The light distribution of the host galaxy is invaluable to providing constraints on the lens model, which in turn provides a theoretical estimation for the time delay associated with each image.
Any variable point source may then be used to measure the time delay.
Since AGN emit unique and stochastic signals, they leave behind temporal fingerprints which have been exploited to determine these time delays at high redshifts~\citep[e.g.][]{Millon20}.
Furthermore, lensed supernovae offer a complimentary target to probe the low redshift Universe~\citep[e.g.][]{Wojtak19, Arendse24}.

Each image in a strongly lensed AGN will emit this unique intrinsic signal.
However, compact objects within the lensing galaxy will modulate the signal due to an effect known as microlensing.
Microlensing affects each image individually, and typically occurs over month to year time scales.
Its impact depends on the lensing configuration of the system~\citep[see][for a review]{Vernardos23}.
The frequency, amplitude, and duration of these microlensing events varies between systems and should be considered whenever strongly lensed sources are analysed~\citep[e.g.][]{Sluse11, Vernardos14, Neira20}.

In a broad sense, microlensing is sensitive to the source size with respect to the average microlens' Einstein radius~\citep{Mosquera11}.
This leads to \emph{differential} microlensing of the various components that make up an AGN~\citep{Wambsganss91, Anguita08, ODowd11, Bate18}.
Microlensing has been successful in placing constraints on multiple size scales of AGN components, from the broad line region (BLR)~\citep{Lewis98, Abajas02, Paic22, Williams21} to optical accretion disc~\citep{Pooley07, Morgan10, Blackburne11, Jimenez12}.
Furthermore, microlensing has the ability to probe the innermost regions of the accretion disc, using both spectral~\citep{Chartas17, Ledvina18} and temporal data~\citep{Abolmasov12a, Mediavilla15b, Tomozeiu18, Best24}.

Fully understanding the structure of AGN is a unique challenge due to two major factors: the cosmological distances involved and the complexity of the signals we observe.
Currently it is believed that most AGN share a similar structure, but the geometry of each component proves difficult to constrain~\citep{Antonucci93, Elvis00}.
The most sensitive instruments can only spatially resolve images of the nearest galactic nuclei.
These include GRAVITY at the Very Large Telescope Interferometer (VLTI), which has resolved the BLR of AGN in the local Universe (z < 0.3) at near-infrared wavelengths~\citep{GRAVITY20b, GRAVITY21, GRAVITY24}, and the Event Horizon Telescope (EHT) which has resolved some of the nearest black holes down to the vicinity of the SMBH~\citep{Horizon19, Horizon22}.
While these methods are powerful, they do not have the ability to be scaled up to the population levels of AGN, especially those at at high redshifts.

Fortunately, the AGN's structure is also encoded within its variability~\citep{Edelson88, Bauer09}.
It is possible to infer model parameters by studying long term light curves across multiple wavelengths.
This has been done in a few cases through reverberation mapping, a technique which assumes that signals propagate through the structure of an AGN, leading to a natural correlation at different wavelengths.
By studying these correlations, reverberation mapping of the continuum has probed the structure of accretion discs~\citep{Mudd18, Jha22, Guo22}, while reverberation mapping of emission lines has provided mass estimates for SMBH~\citep{Peterson04, Peterson14, Grier19, Cackett21, prince22, Kovacevic22}.

In the near future, the Legacy Survey of Space and Time (LSST) is expected to discover and monitor upwards of ~$\mathcal{O} (10^{7})$ AGN and ~$\mathcal{O} (10^{4})$ strongly lensed AGN~\citep{Oguri2010, Yue22}. 
These survey telescopes will not have the spatial resolution to resolve AGN like GRAVITY and the EHT, but instead will observe them with regular monitoring over decade long time scales.
These long term light curves are precisely what is required for probing AGN at the population level.
It is therefore important to understand how particular components of the AGN interact with each other, as opposed to treating them individually.

The accretion disc is believed to produce the optical continuum and has been traditionally modelled with an optically thick and geometrically thin structure where matter accretes onto the black hole~\citep{Bardeen70, ShakuraSunyaev, PageThorne74}. 
This releases gravitational energy and leads to an analytic temperature profile which is then assumed to radiate like a black body~\citep{Planck67}.
However, the effective temperature profiles of accretion discs appear to have significant variance between sources~\citep{Poindexter08, Blackburne15, Munoz16}.
Furthermore, various accretion modes are predicted for different accretion rates such as slim discs, thick discs, and advection dominated accretion flows~\citep[see][for a review]{Abramowicz13}.

For any accretion disc model, the gravitational potential of the SMBH influences how we perceive the disc. 
The innermost regions of the accretion disc will experience the most intense distortions by GR ~\citep{Rees84}.
It has been shown that the distribution of radiation from the thin disc is expected to be altered by GR, especially when the disc is highly inclined with respect to the observer~\citep{NovikovThorne73, PageThorne74, Jaroszynski92}.
The gravity of the SMBH causes both light bending and redshifting of the emitted radiation~\citep{Abolmasov12a, Mediavilla15b, Chartas17, Ledvina18, Kammoun19, Best24}.
Therefore, the interplay between GR and the models building up an accretion disc is an important step in producing accurate surface brightness distributions.

The BLR is a well known component associated with emission lines which are broadened by thousands of km s$^{-1}$, and plays a significant role in our understanding of type 1 AGN~\citep{Schneider90, Peterson93, Murray95, Proga00, Elvis00, Sluse07, ODowd11, bentz13}.
Traditionally, reverberation mapping of certain emission lines in the BLR has provided mass estimates for the SMBH~\citep{Peterson04, Peterson14, Grier19, Cackett21}.
This assumes the continuum component of the AGN drives the variability of the emission lines.
Often the BLR is modelled as a virialized disc with a characteristic radius which leads to a simple mass estimate.
However, the explicit geometry can introduce uncertainties in this mass estimate up to a factor of $\mathcal{O}(10)$.
This is typically accounted for by a virial factor which is inclination and model dependent~\citep[e.g.][]{Mejia18}.

It is possible to use comprehensive and versatile models to understand the BLR.
In order to reverberate, the BLR must absorb radiation at specific wavelengths prior to fluorescence~\citep{Mccrea36, Rottenberg52, Murray95}.
These processes will modify the observed spectra of AGN based on the relative strengths of emission lines~\citep{Blandford82}.
Non-trivial geometries may impart their signature on these spectral features of the BLR~\citep{Waters16, Yong17, Hutsemekers21, Ng23}.
By studying the properties of these spectral lines, we can gain information about the kinematics and geometry of the BLR.  

A dusty torus is believed to exist beyond the accretion disc and the BLR.
Its impact on AGN spectra is two-fold: to provide obscuration leading to the dichotomy between type 1 and type 2 AGN~\citep[e.g.][]{Antonucci93} and be responsible for excess mid-infrared emission observed across many AGN~\citep{Dullemond05, Nenkova08}.
The torus is also known to reverberate at very long timescales with respect to other AGN components~\citep{Pier92, Cackett21}.
The composition of the torus is likely molecular dust composed of graphite and silicates, both of which have been studied extensively for their extinction and reddening properties~\citep{Draine84, Ramos09}.
Obscuration by the torus is an important feature and should be accounted for, especially when modelling AGN at high inclination angles.

With multiple pieces of the AGN model in place, the variability should reflect each component.
There have been analyses of AGN by treating optical AGN variability as a damped random walk (DRW) or other higher order continuous auto-regressive moving average (CARMA) processes~\citep{Kelly09, Kelly14, Moreno19, Yu22}.
CARMA parameters and AGN properties are often found to be correlated~\citep{macleod10, Suberlak21, Yuk23}.
However, there is also evidence against these models fully accounting for AGN variability~\citep[e.g.][]{Kasliwal15}.
There are alternative methods for modelling AGN variability, such as using machine learning algorithms consisting of a recurrent auto-encoders~\citep{Tachibana20}.
These make no prior assumptions on the AGN model and have been successfully used to model AGN variability.
Other model dependent machine learning algorithms such as latent stochastic differential equation networks have the ability to simultaneously model AGN variability as well as predict model parameters~\citep{Fagin24a}.
Furthermore, magneto-hydrodynamic simulations may be used to simulate variability via radiative reprocessing in the accretion disc~\citep[e.g.][]{Secunda24}.

One simple and analytic technique for modelling AGN variability is the lamp-post model, where a compact X-ray emitting corona irradiates the accretion disc~\citep{Collin03, Sergeev05}.
These photons are believed to originate from seed photons from the inner accretion disc that become Compton scattered to X-ray energy levels within a compact region near the SMBH~\citep{Shapiro76, Haardt91, Matt92, Markoff05}.
The disc is then assumed to thermally reprocess this radiation after a time lag dominated by the light travel time~\citep{Cackett07, Kammoun19, Chan20, Jha22, Guo22, chan24}.
This leads to optical variability, which is correlated with the X-ray signal.
Some studies have shown that the X-ray Power Spectrum Density (PSD) typically follows a bending power law~\citep{Edelson99, Papadakis04}.
When the accretion disc reprocesses this signal, the PSD is affected due to the relatively large size of the accretion disc.

The reprocessing of intrinsic variability within the lamp-post model is intricately tied to the temperature profile.
There is an important relationship between properties of the accretion disc and other components such as the SMBH, the corona, and the BLR.
Parameters like the accretion rate and the SMBH mass impact the reprocessing of X-ray photons into the stochastic optical emission~\citep{Kammoun21}.
These are important to keep constant throughout each AGN component to remain consistent.
Some work has been done on exploring these models for various temperature profiles ~\citep[e.g.][]{chan24} and applying these models directly in neural networks through auto-differentiable framework ~\citep[e.g.][]{Fagin24b}.

Intrinsic variability continues beyond the accretion disc to the BLR~\citep{Eracleous, Paic22, Rosborough23}.
The BLR may be responsible for the diffuse continuum in addition to its characteristic broad emission lines.
The diffuse continuum is believed to cause the discrepancy of accretion discs appearing too large with respect to standard theory~\citep{Korista01, Chelouche19, Korista19, Netzer22}, while reverberating broad emission lines are an important estimator of the SMBH mass~\citep[e.g.][]{Peterson14}.
A self-consistent model for intrinsic variability in AGN requires a connection across all size scales, from the corona to the accretion disc and the BLR.
We expect that different structures will naturally have different reverberation properties.
The shape and geometry of each component will determine how the signal is reprocessed.

When studying strongly lensed AGN, and in particular lensed quasars, it is often desirable to separate the intrinsic variability from microlensing.
However, the observed variability of a system will not simply be the sum of the components.
Differential microlensing shows us the importance of spatial separation, while models for intrinsic variability are derived from the propagation of signals.
The sum of variabilities represent the first order stochastic signal of the system, but higher order effects have only recently become a topic of consideration~\citep[e.g. the microlensing time delay, ][]{Tie18, Chan21}.
Disentangling these contributions is crucial for fully understanding our underlying model assumptions and is an important consideration for the analysis of observed light curves of these systems.

In this paper, we introduce \texttt{Amoeba}: an AGN Model of Optical Emissions Beyond steady-state Accretion discs. 
This code aims to generate mock light curves for AGN such as those which will be discovered and monitored by wide-field surveys such as LSST hosted at the Vera Rubin Observatory.
\texttt{Amoeba} simultaneously creates accretion disc surface flux distributions, accretion disc responses to stimuli, and BLR simulations in order to treat intrinsic variability throughout the AGN model.
The methods used are flexible enough to allow for arbitrary accretion disc profiles and BLR geometries to facilitate joining together various models.
These reverberating AGN components may then be used to consistently simulate signals from both lensed and non-lensed AGN while still incorporating relativistic effects due to the central SMBH.

In Section~\ref{AccretiondiscSection} we introduce the accretion disc model and describe its ability to generate surface flux distributions for arbitrary thermal profiles while incorporating relativistic effects.
Section~\ref{BLRsection} details how the BLR is modelled for any axisymmetric region and provides an example of a bi-conical BLR.
By attributing spectral broadening to line-of-sight velocities, we show how spatially distinct regions of the BLR may contribute to different optical bands.
In Section~\ref{TorusSection} we describe how an obscuring torus may be simulated alongside its impact on spectral features as a function of inclination.
Section~\ref{VariabilitySection} describes the intrinsic variability models within \texttt{Amoeba} and the self-consistent propagation of a signal throughout the AGN from the corona through the BLR with the inclusion of increased inter-band continuum time lags due to the diffuse continuum.
Various transfer functions are calculated to represent the response of accretion disc to the corona and the response of the BLR to the accretion disc. 
Section~\ref{MicrolensingSection} presents the simulation of microlensing light curves for both the continuum and the BLR.
In Section~\ref{FullVariabilitySection} we describe the interplay between variability sources, discuss microlensed transfer functions, and create intricate light curves with mock observations which reflect the multiple components of the AGN model.
Lastly, Section~\ref{ConclusionSection} concludes with some use cases of \texttt{Amoeba} and describes some future planned modules.
Examples assume a flat $\Lambda$CDM cosmology with H$_{0}$ = 70 km s$^{-1}$ Mpc$^{-1}$ and $\Omega_{\rm{M}} = 0.3$.

\section{Accretion Disc Modelling}
\label{AccretiondiscSection}
AGN accretion discs have been modelled in varying levels of complexity~\citep{Salpeter64, Abramowicz88}. 
Simple cases such as Gaussian flux distributions are typically used when the structure of the disc has little effect on the desired output~\citep[e.g. long-term microlensing analysis][]{Grieger88, Jimenez14, Bate18}.
A more physical model is the geometrically thin and optically thick Shakura-Sunyaev (SS) accretion disc, colloquially the thin disc, which is an appropriate baseline for AGN accreting at sub-Eddington rates~\citep{ShakuraSunyaev}.
The Novikov-Thorne (NT) profile builds upon this with relativistic corrections~\citep{PageThorne74}.
Accretion discs accreting at higher rates near or above the Eddington limit require more efficient accretion flows such as the slim disc~\citep{Abramowicz13, Czerny19}, while very low accretion rates may follow advection-dominated accretion flows~\citep{Narayan95, Fabian95}~\citep[see][for a review]{Yuan14}. 
The accretion disc may also be modelled using state-of-the-art general relativistic magneto-hydrodynamic (GR-MHD) simulations, which allows for the accretion disc to evolve with time following GR-MHD equations~\citep[e.g.][]{Kaaz23, Musoke23, Secunda24}.
However, even with the improvements to GR-MHD simulations over the past few years, they require long computation times to evolve into a steady state for only a single set of model parameters~\citep[e.g.][]{Liska22}.
Due to this, we aim to use a time-averaged thermal profile akin to those in~\citet{Fagin24a} and~\citet{chan24}, then include relativistic effects such as light bending, relativistic beaming, and relativistic Doppler shifting.

\subsection{Effective Temperature}

The time-averaged thin disc temperature profile is appealing for its physically derived and analytic form.
The resulting temperature profile can be described as in~\citep{ShakuraSunyaev}:
\begin{equation}
  \label{Thindisc}
  T^{4}_{\rm{SS}}(R) = \frac{GM_{\rm{BH}}\dot{M}\left(1 - \sqrt{r_{\rm{in}} / R}\right)}{8\pi \sigma_{\rm{sb}} R^{3}} ,
\end{equation}
where $G$ is the gravitational constant, $M_{\rm{BH}}$ is the mass of the central black hole, $\dot{M}$ is the accretion rate, $r_{\rm{in}}$ is the inner radius of the accretion disc, and $\sigma_{\rm{sb}}$ is the Stefan-Boltzmann constant. 
At large radii, this leads to the often cited $T \propto R^{-3/4}$.

It is common to express the accretion rate in terms of the Eddington ratio--the accretion rate required for the accretion disc to radiate at the Eddington limit.
The Eddington luminosity is defined by balancing the gravitational force against the radiation pressure of accreted protons. 
This leads to $L_{\rm{Edd}} = 4 \pi G M_{\rm{BH}} m_{\rm{p}} c / \sigma_{\rm{T}}$ where $m_{\rm{p}}$ and $\sigma_{\rm{T}}$ are the mass of the proton and Thompson cross section of the electron, respectively.
The bolometric luminosity of the disc is the total energy emitted through the release of gravitational energy, defined as $L_{\rm{Bol}} = G \dot{M} \eta / c^{2}$ where $\eta$ is an efficiency factor for the release of gravitational potential energy and is related to the spin of the SMBH.
The Eddington ratio is then the accretion rate with respect to the accretion rate required to radiate at the Eddington luminosity.

Within the Kerr metric~\citep{Kerr63}, the inner limit of the thin disc is taken to be the Innermost Stable Circular Orbit (ISCO) labelled as $r_{\rm{ISCO}}$.
As the SMBH's spin approaches the limiting values of 0, 1, and $-1$, the ISCO radius approaches 6, 1, and 9 gravitational radii ($r_{\rm{g}} \equiv GM_{\rm{BH}}/c^2$).
Beyond the Kerr metric, it has been shown that electrically charged SMBHs may affect $r_{\rm{ISCO}}$, with the ability to mimic a wide range of spins~\citep{Zajacek18, Zajacek19, Tursunov20}.
We do not consider the possibility of charged black holes within \texttt{Amoeba}, though this may be a topic of a future extension.

\begin{figure}
    \centering
    \includegraphics[width=0.47\textwidth]{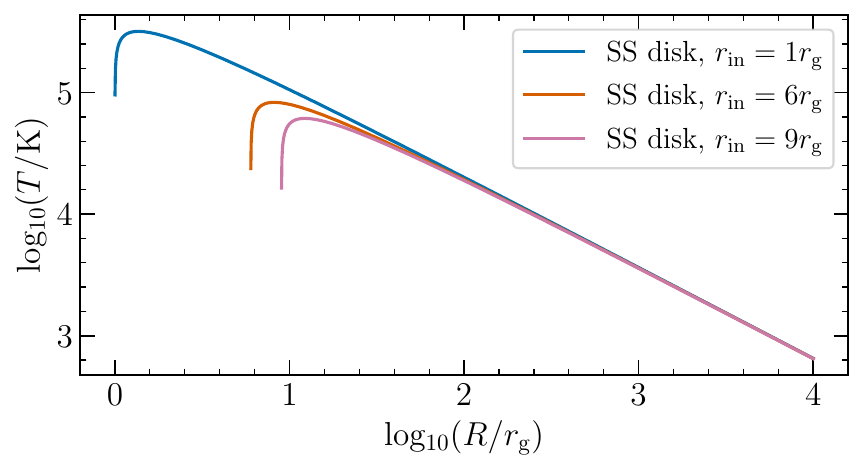}
    \caption{Temperature profiles for the SS disc model assuming $r_{\rm{ISCO}}$ = 6, 1, and 9 $r_{\rm{g}}$. Other parameters include $M_{\rm{BH}} = 10^{8} M_{\odot}, ~r_{\rm{in}} = r_{\rm{ISCO}}$, and an Eddington ratio of 0.15. Temperatures increase as the inner limit is reduced. All temperature profiles converge for large radii.}
    \label{IMGRinTemps}
\end{figure}

Since $r_{\rm{ISCO}}$ is spin dependent, defining $r_{\rm{in}} = r_{\rm{ISCO}}$ leads to a change of the temperature profile on the smallest scales near the black hole according to Equation (\ref{Thindisc}), though the outer regions are relatively insensitive to this choice. 
Fig.~\ref{IMGRinTemps} illustrates that prograde accretion discs may have an exceptionally hot inner region in the thin disc model due to $r_{\rm{ISCO}}$ becoming extremely small. 
Despite inner regions being significantly hotter than other Schwarzschild or retrograde Kerr discs, the inner region does not contribute much to the overall optical flux due to the very small emission region.
However, this region is important for UV/X-ray emission or cases where the inner accretion disc lays on a caustic fold (see Section~\ref{MicrolensingSection}).

Within \texttt{Amoeba}, we aim to allow for a flexible environment to incorporate as many models as possible.
A flexible and physically motivated temperature profile which converges to the thin disc is defined in Appendix~\ref{appendix_temp_profile} which is comparable to the model presented in~\citet{chan24}. 
However, we note that any 1 or 2 dimensional effective temperature profile may be used to generate black body radiation from the accretion disc.

\subsection{Surface Flux Distribution}
\label{ObservationalEffects}

\begin{figure*}
    \includegraphics[width=0.97\textwidth]{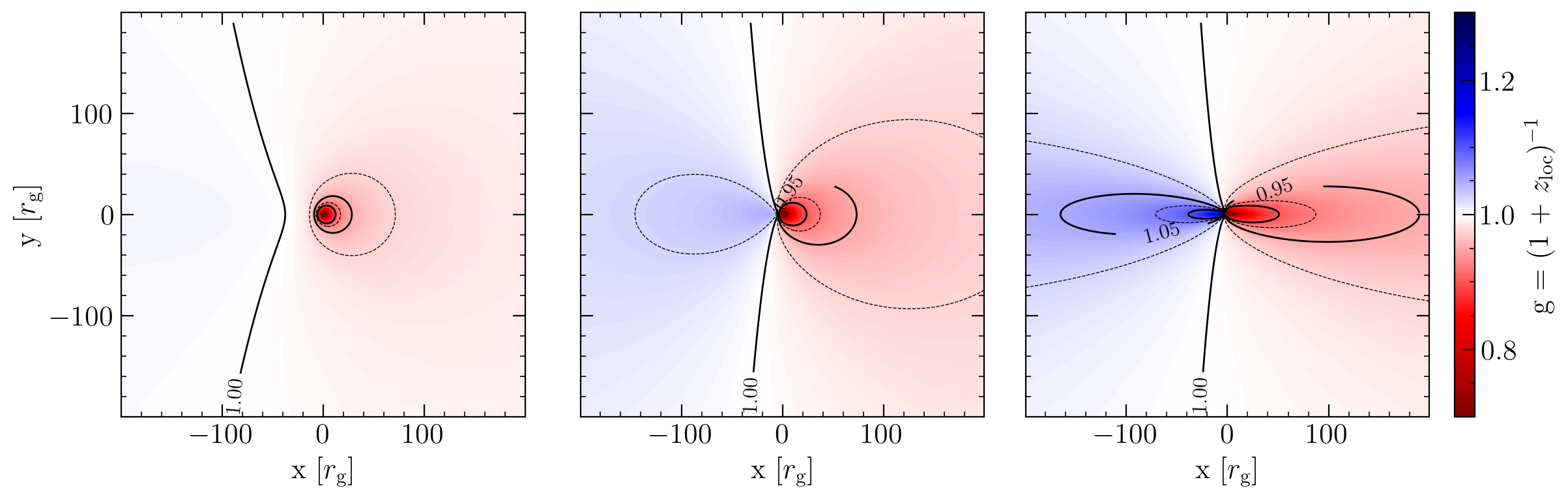}
    \caption{Local redshift factor for Schwarzschild accretion discs with 10$\degree$, 30$\degree$, and 70$\degree$ inclination from left to right. As inclination increases, the redshift factors increase/decrease for the approaching/receding sides. Selected contour levels are highlighted for clarity and dotted contours are evenly spaced between solid contours at $g = 0.925, 0.975, 1.025, 1.075$.}
    \label{FigLocalRedshifts}
\end{figure*}

\begin{figure}
    \centering
    \includegraphics[width=0.47\textwidth]{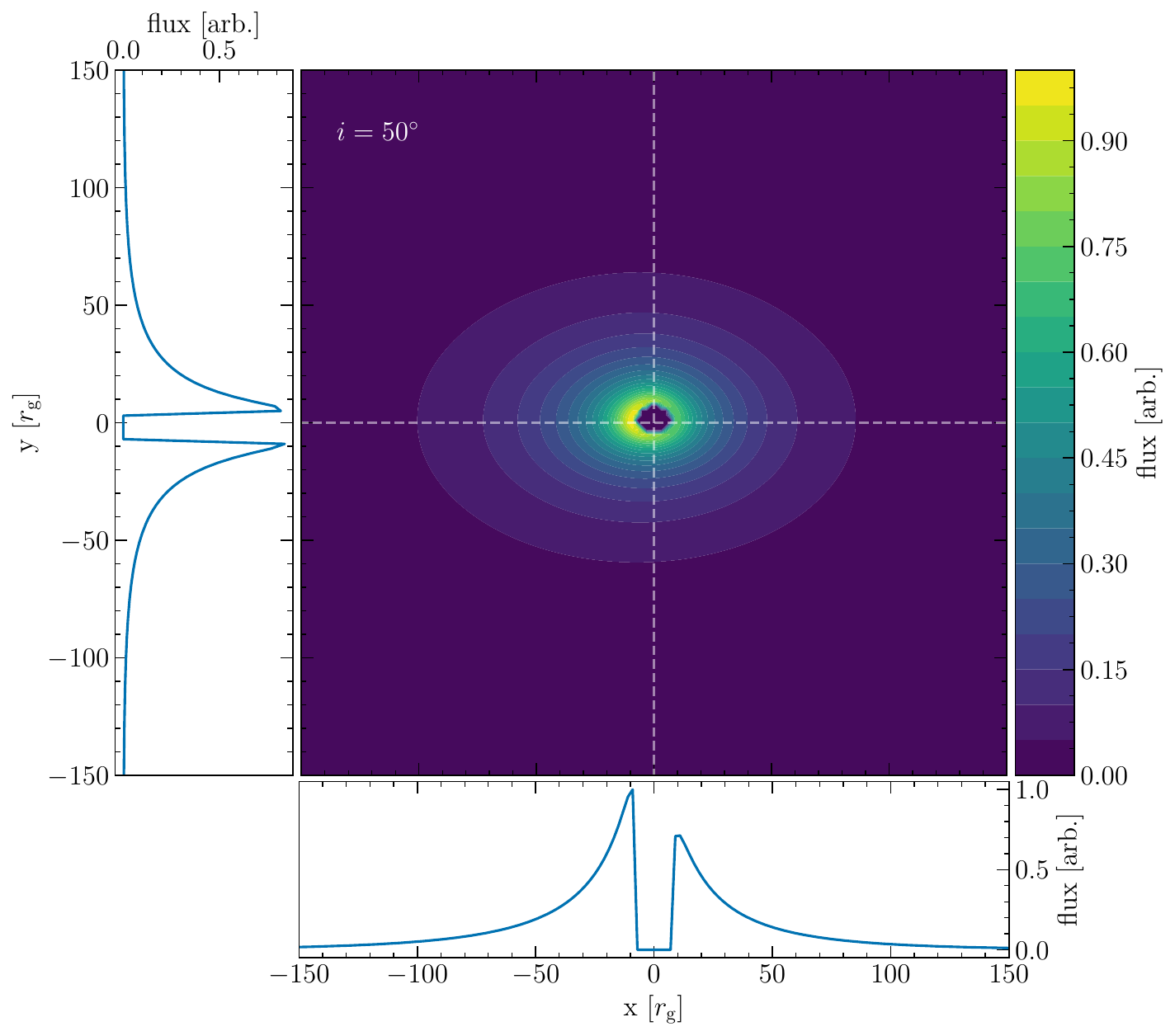}
    \caption{Example surface flux distribution for a thin disc viewed at inclination $50\degree$ emitting at 900 nm in the rest frame. Cross sectional profiles are taken to emphasize asymmetry in the accretion disc's surface flux distribution.}
    \label{FigSurfaceBrightness}
\end{figure}

The thermal profile of the accretion disc cannot be observed directly.
Instead, it is the total flux that gets shifted into some wavelength range which is observed.
\texttt{Amoeba} is designed to be semi-analytic, support microlensing, and support obscuration, so these flux distributions must be calculated.
To produce surface flux distributions from temperature distributions, we assume the accreted material radiates in its own frame of reference as a black body, where the spectral radiance is defined as~\citep{Planck67}:
\begin{equation}
    \label{Planck}
    B(\lambda, T) = \frac{2hc^{2}}{\lambda^{5}} \left( \frac{1}{e^{hc/\lambda k_{\rm{B}} T} - 1} \right)
\end{equation}
where $h$ is Planck's constant, $\lambda$ is the rest wavelength, $k_{\rm{B}}$ is the Boltzmann constant, and $T$ is the temperature.
$B(\lambda, T)$ should be integrated over the accretion disc's surface and a range of wavelengths to obtain the flux.

Equation (\ref{Planck}) is evaluated at wavelengths $\lambda_{\rm{rest}}$ which will become redshifted (blueshifted) to the simulated observed wavelength $\lambda_{\rm{obs}}$.
This redshifting (blueshifting) incorporates the relativistic effects of Doppler beaming, Doppler shifting, and gravitational redshift of the photons emitted in the rest frame of the accretion disc, as well as cosmological redshift.

In order to produce the accretion disc projection, \texttt{Amoeba} uses \texttt{Sim5}~\citep{Bursa17} to trace photons from the observer to the accretion disc. 
Photon paths are deflected due to the local gravitational field of the central black hole, independent of a gravitational lens located between the observer and the source.
Increasing the inclination amplifies these effects. 

Cosmological redshift affects every aspect of the simulation in both terms of wavelength scaling and time dilation.
GR ray-tracing codes calculate relativistic redshifts that arise due to light travelling through the metric surrounding the black hole which we label $z_{\rm{loc}}$.
Furthermore, we define the relativistic redshift factor $g \equiv 1 / (1 + z_{\rm{loc}})$ as the impact local redshifting has on observed photons.

Fig.~\ref{FigLocalRedshifts} illustrates the evolution of the local redshift factor $g$ for inclined accretion discs as calculated by \texttt{Sim5}. 
In each case, the left side is approaching the observer and is blueshifted by Doppler shifting, while the right side is receding and will be redshifted.
Gravitational redshift affects both sides equally, and is most prominent near the central black hole where the gravitational potential is the strongest. 
Gravitational redshifting is the leading redshifting component at 10$\degree$ inclination, while Doppler shifting is the leading component at 70$\degree$.

This redshift adjusts the observed intensity of light as it leaves the Kerr metric of the central black hole.
The observable intensity of radiation follows $I_{\rm{obs}} = I_{\rm{emit}}g^{4}$ ~\citep{LINDQUIST66, Ellis71, Cunningham73}.
$I_{\rm{obs}}$ is the intensity as seen by the observer, and $I_{\rm{emit}}$ is the intensity emitted by the source.
In the absence of a strong gravitational well or high line-of-sight velocities, this factor reduces to unity everywhere.

Fig.~\ref{FigSurfaceBrightness} represents a surface flux distribution for a thin disc temperature profile considering relativistic effects. 
The flux has been normalized and the accretion disc appears brighter on the left, approaching side. 
The redshift from the gravitational potential effects the locally emitted radiation which gets shifted into the observed wavelength. 
Therefore, the observed surface flux distribution has contributions from a variety of locally emitted wavelengths calculated at the pixel level.

\section{Broad Line Region Modelling}
\label{BLRsection}
The Broad Line Region (BLR) is observed as a series of emission and absorption lines, which contain contributions from various ionic species across the electromagnetic spectrum.
While most AGN produce broad emission lines, only a small subset ($\sim$ 1 - 10 per cent) exhibit absorption lines known as broad absorption line (BAL) AGN~\citep{Gibson09}.
The emission lines are excited by the AGN's continuum and fluoresce at velocity shifted wavelengths which leads to a broadened spectrum.
Important emission lines include low-ionization lines (e.g. H$\alpha$, H$\beta$, Mg II)~\citep{Kaspi01, McLure02, Sluse07, Zu11, bentz13, grier17, Mejia18, Hutsemekers19, Guo22} and high-ionization lines (e.g. C IV)~\citep{Vestergaard02, Vestergaard06, Williams21}.
The reverberation of these lines is one of the primary methods for probing the AGN's geometry and kinematics, and low-ionization lines provide an important estimator for the mass of the AGN's black hole~\citep{Blandford82, Kaspi07, Peterson14, Grier19, Hoormann19, Czerny19b, Zajacek20, Homayouni20, Yu21}.
Often the emission lines are fit as a sum of functions representing emission templates that consider redshifting (blueshifting) and other broadening effects.

Within this section, we discuss the presence of BLR's emission to LSST-like filters.
The filters are defined as top-hat functions, spanning the wavelength ranges with transmission greater than 1 per cent.
For simplicity, we do not consider the specifics of the filter beyond minimum and maximum wavelengths which we take to be for the $z$ and $y$ filters to be [802, 936] and [912, 1075] nm, respectively.
Within this work, the BLR is modelled by exploiting cylindrical symmetry. 
As such, we use coordinates ($R, \phi, Z$) to represent cylindrical coordinates as opposed to ($r, \phi, \theta$) for spherical coordinates.

\subsection{BLR Wind Region}
\label{BLRWindsection}
\texttt{Amoeba} has the ability to model the BLR for any geometry with cylindrical symmetry.
For outflows which are driven by continuum radiation or line-driving effects, the BLR may be defined by a set of streamlines traced onto a set of $R, Z$ coordinates.
The velocity along these streamlines may be defined as in~\citet{Yong17}:
\begin{equation}
    v_{\rm{pol}}(l) = v_{0} + (v_{\infty} - v_{0}) \left( \frac{(l/R_{v})^{\alpha_{\rm{w}}}}{(l/R_{v})^{\alpha_{\rm{w}}} + 1} \right)
    \label{PoloidalVelocityEqu}
\end{equation}
where $l$ is the poloidal distance along a streamline, $v_{0} \text{ and } v_{\infty}$ are the initial and asymptotic velocities respectively, $R_{v}$ is the characteristic distance at which the streamline approaches $v_{\infty}$, and $\alpha_{\rm{w}}$ is a power law index which determines how rapidly $v_{\infty}$ is approached.
The poloidal direction is defined as the linear direction of the streamline's motion in the ($R, Z$) plane.
The angular velocity $v_{\phi}$ is assumed to remain Keplerian around the axis of cylindrical symmetry.

\texttt{Amoeba} uses streamlines as the boundaries to the broad line flow and interpolates between for intermediate values.
Any number of wind regions may be added together to define complex wind features. 
The density of the wind at any region is calculated assuming conservation of mass.
The line emitting species is assumed to be confined to this region.

Line emitting species emit photons at particular wavelengths when an electron relaxes to a lower energy state.
In doing so, the electron emits a photon defined by the energy gap between quantum states.
This process can occur in two different ways: An electron may be in an excited energy level and drop down to a lower energy state through relaxation, or an ionized species may capture a free electron through recombination~\citep{Peterson01}.
These are also known as bound-bound and bound-free interactions, respectively.
While these processes may be similar (e.g. an energetic electron emits a photon by lowering its energy state), they depend on slightly different parameters.

Bound-bound emission lines do not have to be ionized as the photon comes from transition in electron energy states (e.g. from an excited state to a rest state).
These emission lines are quantized due to the electrons' discrete initial and final states.
On the other hand, bound-free emissions can lead to what is known as the diffuse continuum, as the electron's initial state is not quantized~\citep{Lawther18, Korista19, Chelouche19}.

In both cases, the fraction of line emitting particles that are ionized or excited to the line-emitting state is related to the ionization potential through the Saha ionization equation~\citep{Saha1921}.
We do not repeat this equation or directly use it, as it becomes unwieldy once more than one ionization state is considered and should be handled with photo-ionization codes~\citep[e.g. \texttt{CLOUDY}][]{Ferland17}.
What is important is that regions of high ionization potential (e.g. near the SMBH and unshielded regions above the accretion disc within the ionization cone) will favour higher ionization states compared to the usual line emitting states, and regions heavily shielded or obscured will not quite reach the energetic level required to effectively emit the line.
However, there will be some region in which the ionization state will be just right to facilitate strong line emission.
This may be modelled using the local optimally emitting cloud (LOC) model~\citep{Baldwin95}.
The LOC region can then be defined within \texttt{Amoeba} to target optimally emitting regions of the BLR to be used in conjunction with the accretion disc both for the flux distribution of the BLR and reverberation of emission lines.

\subsection{Sample BLR Contamination}
\label{SampleBLRContaminationSubsection}

To illustrate the BLR, we consider a simulated emission line of 486 nm for an AGN positioned at $z_{\rm{s}} = 1$.
We set $i = 45 \degree$ and simulate emission as a contaminant for the LSST $y$ and $z$ filters. 
We define a fiducial BLR geometry as a bi-conical outflow bounded by two streamlines with parameters defined in Table~\ref{TableStreamlineParams}.
All angles are with respect to the normal to the accretion disc plane, such that $\theta_{\rm{SL}, x} = 0\degree$ would represent material being ejected normal to the accretion disc.

Within the wind region, the emission from the BLR is assumed to follow the LOC model.
We make the assumption that the BLR preferentially emits from a spherical shell region defined by a Gaussian centred at 300 $r_{\rm{g}}$ with width of 50 $r_{\rm{g}}$ and stress this example is only for illustrative purposes.

\begin{figure}
    \centering
    \includegraphics[width=0.47\textwidth]{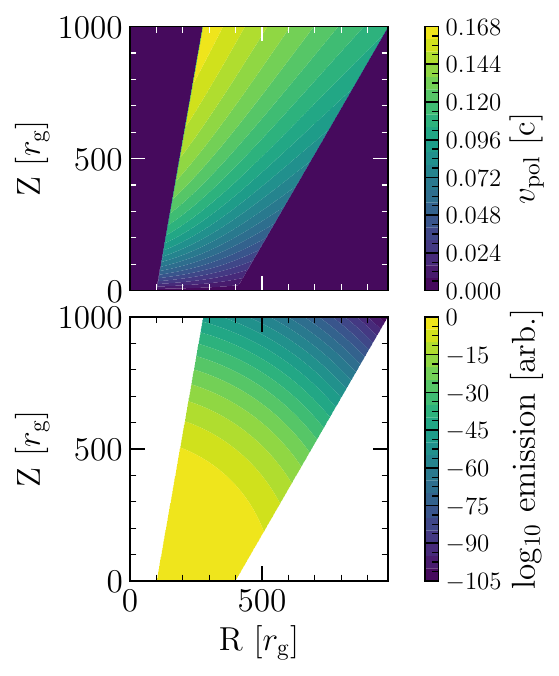}
    \caption{Top: Distribution of poloidal velocities in the R-Z plane of the BLR. The streamlines described in Table ~\ref{TableStreamlineParams} define the boundaries of the BLR. Bottom: Effective weighting of the BLR emission in the R-Z plane. This weighting depends on the relative particle density and the assumed optimal emission region defined by a Gaussian function centred at 300 $r_{\rm{g}}$ with width 50 $r_{\rm{g}}$.}
    \label{blr_rz_geometry}
\end{figure}

Fig.~\ref{blr_rz_geometry} illustrates the geometry of the example BLR.
The upper panel describes the spatial distribution of poloidal velocities.
As height increases from the plane of the accretion disc, particles are assumed to be accelerated outward via line-driving effects following Equation (\ref{PoloidalVelocityEqu}).
The lower panel represents the effective emission of each location in the R-Z plane.
As the BLR accelerates, the particle density decreases to conserve particle flux. 
The density is then weighted by the assumed optimal emitting region, leading to preferential emission within $\sim 500~r_{\rm{g}}$ from the SMBH.
We note that we do not consider shielding by dust or the inner regions of the BLR at this time.

The BLR then is assumed to contaminate certain filters where the emission line is shifted into.
\texttt{Amoeba} considers the shift of the emission line to be due to the line-of-sight velocity components of the BLR.
Contributing factors include the Keplerian motion, poloidal motion, and cosmological redshift.
Pressure broadening and isotopes are not considered, as these are negligible in comparison to other effects.

\begin{figure}
    \centering
    \includegraphics[width=0.47\textwidth]{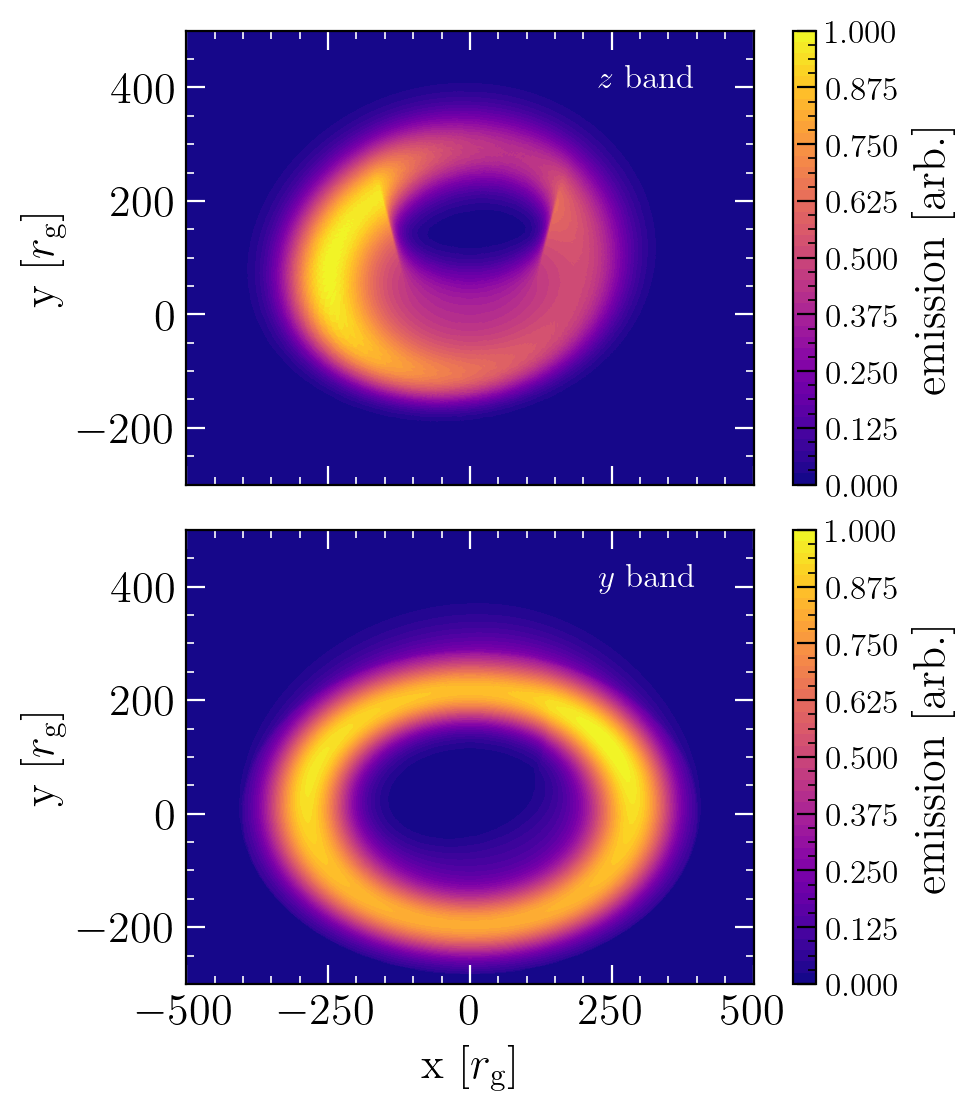}
    \caption{Projections of the BLR contaminant to LSST-like filters. The top panel represents the contribution to the $z$ filter, while the lower panel represents the contribution to the $y$ filter. Doppler shifting projects a component of the BLR into the $z$ filter.}
    \label{ProjectedBLR}
\end{figure}

Fig.~\ref{ProjectedBLR} represents the projected BLR into both the $z$ and $y$ LSST-like filters defined as top-hat functions within the ranges [802, 936] and [912, 1075] nm, respectively. 
The asymmetry arises from the required Doppler shift to bring the rest frame emission line into wavelengths associated with the $z$ filter.
The consequences of this asymmetric distribution will be explored with respect to reverberation mapping and microlensing in Sections~\ref{BLRTransferFunctionSection} and~\ref{BLRMicrolensingSection}, respectively.
This emission line model does not project into the other LSST-like filters, so line contamination in this case is restricted to the $y$ and $z$ bands.

\section{Torus Modelling}
\label{TorusSection}
AGN are believed to host a dusty torus due to the spectral differences observed between type 1 and type 2 AGN.
The inner edge of tori are believed to consist of molecular dust including graphite and silicates~\citep{Draine84}.
We highlight these particular species due to their relatively high sublimation temperatures ranging from 1500 - 1800 K~\citep{Ramos09, Gonzalez19}.
Other studies have shown that the BLR may typically exist at radii relative to the critical temperature for Hydrogen dust around 1000 K~\citep{Czerny11}.
Within this section, we take the start of the torus to be the highest temperature of these ranges at 1800 K, and only look at the attenuation of the accretion disc's optical radiation.
In reality, the torus is likely a more complex object and can be modelled many ways.

\begin{figure}
    \centering
    \includegraphics[width=0.47\textwidth]{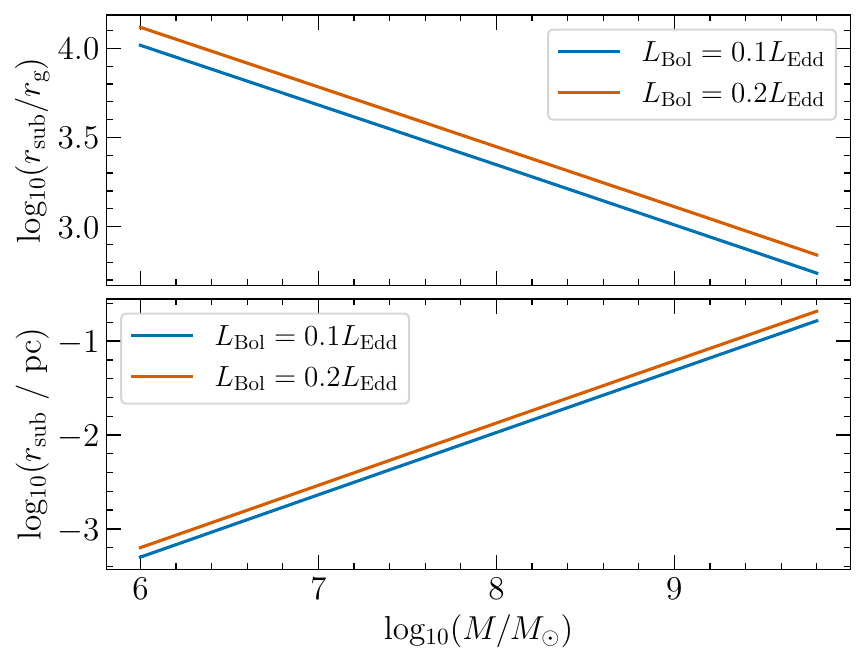}
    \caption{Sublimation radius as a function of SMBH mass. Two accretion rates are chosen such that $L_{\rm{Bol}}/ L_{\rm{Edd}} = $ 0.1 and 0.2.}
    \label{Fig_sublimation_radius}
\end{figure}

Fig.~\ref{Fig_sublimation_radius} illustrates the sublimation radius $r_{\rm{sub}}$ of the torus as a function of mass within this model for two different choices of accretion rates.
The sublimation radius is taken to be the radius where the temperature of the accretion disc drops below 1800 K~\citep[e.g.][]{Gonzalez19}.
The upper panel represents the sublimation radius in terms of $r_{\rm{g}}$, while the lower panel is scaled to units of pc.

The obscuring torus may be defined as an axi-symmetric region similarly to the BLR.
\texttt{Amoeba} projects column densities onto the source plane in order to determine attenuation based on an absorption profile.
Treating the torus as a simple obscurer will cause colour dependence due to both the spatial flux distributions of the accretion disc and the absorption profile of the dust.

Attenuation of light by a medium may be described as:
\begin{equation}
    I_{\rm{obs}} = I_{\rm{emit}} e^{-\rho_{\rm{col}} / \tau_{\rm{abs}}(\lambda)} ,
\end{equation}
where $I_{\rm{obs}} \text{ and } I_{\rm{emit}}$ are the observed and emitted intensities respectively, $\rho_{\rm{col}}$ is the column density of particles, and $\tau_{\rm{abs}}$ is the wavelength dependent absorption coefficient of the material (not to be confused with time lags).
Wavelength dependent attenuation is an important feature of dust and the torus is likely composed of graphite and silicates~\citep[e.g.][]{Ramos09, Feltre12, Gonzalez19}.
This attenuation model allows for a natural relationship between absorption and column density.
Our implementation allows us to model non-uniform densities as expected in a dusty torus, which are believed to typically decay as a power law and vary with inclination~\citep[e.g.][]{Nenkova08}.

\begin{figure}
    \centering
    \includegraphics[width=0.47\textwidth]{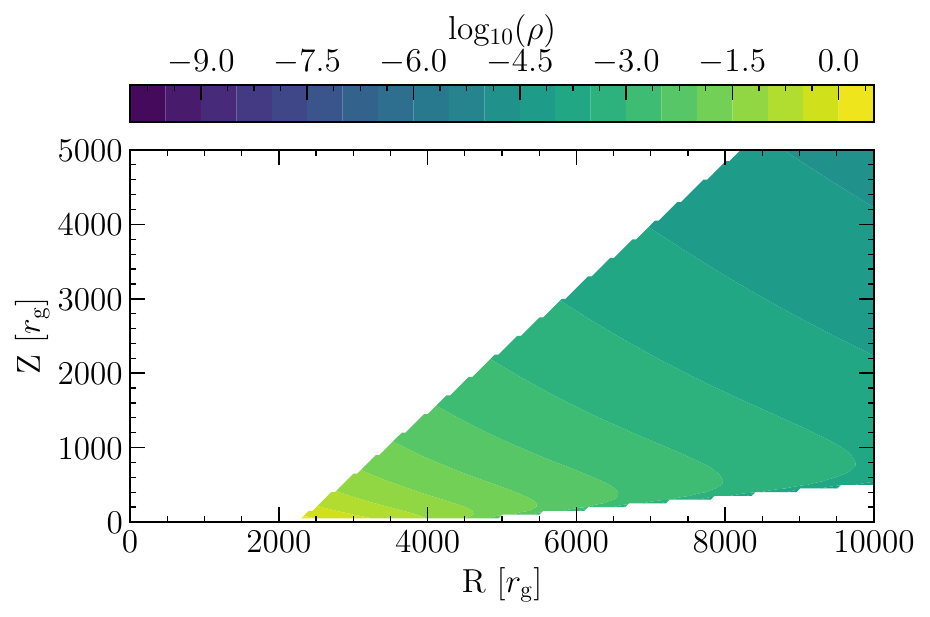}
    \caption{Particle density plot of the dusty torus in the R-Z plane.}
    \label{FigTorusDensity}
\end{figure}

Fig.~\ref{FigTorusDensity} represents the density of a model torus in the R-Z plane, with half-opening angle $40\degree$ and mean density assumed to decay as $r^{-2}$.
The SMBH is assumed to have $M_{\rm{BH}} = 10^{8} M_{\odot}$ and an accretion rate leading to an Eddington ratio of 0.1.
We define the inner radius to be the point where our accretion disc's temperature reaches the dust sublimation temperature $T_{\rm{sub}}$, which we take to be 1800 K, as defined in Fig.~\ref{Fig_sublimation_radius}.
For this accretion disc, the dust sublimation radius is approximately 2200 $r_{\rm{g}}$ and note that the accretion disc may extend beyond this point.
To illustrate the colour dependence this can have on the accretion disc, we focus on wavelengths corresponding to LSST $g$ and $y$ bands within this section.
For simplicity, we define the dust's attenuation to be twice as strong for the $g$ band compared to the $y$ band.

\begin{figure}
    \centering
    \includegraphics[width=0.47\textwidth]{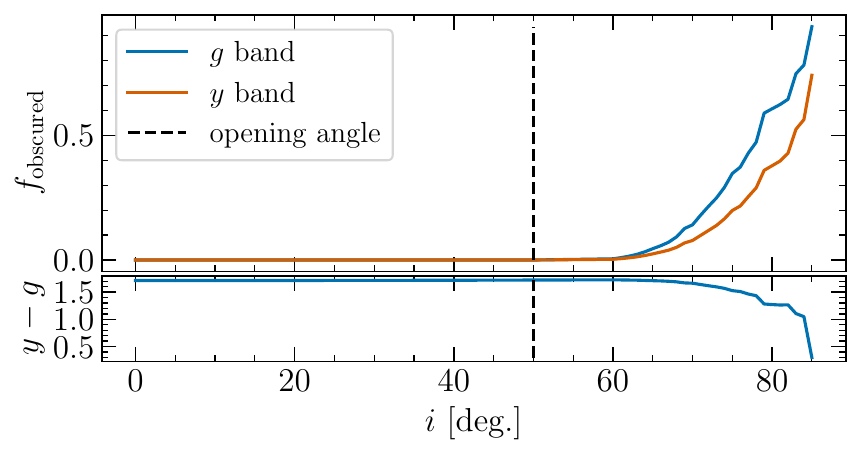}
    \caption{Top: Obscuration factor $f_{\rm{obscured}}$ as defined in Equation (\ref{ObscurationFactorEquation}). Bottom: Magnitude colour difference between $y$ and $g$ LSST-like bands.}
    \label{FigObscurationByTorus} 
\end{figure}

We define the obscuration factor of the accretion disc by the torus as:
\begin{equation}
\label{ObscurationFactorEquation}
    f_{\rm{obscured}}(\lambda, i) \equiv 1 - F_{\rm{obscured}}(\lambda, i) / F_{\rm{emit}}(\lambda, i) ,
\end{equation}
where $F_{\rm{emit}}(\lambda, i) \text{ and } F_{\rm{obscured}}(\lambda, i)$ are the integrated flux distributions of the accretion disc before and after obscuration, respectively.
The integrated fluxes are numerically calculated over all relevant wavelengths.
A value of $f_{\rm{obscured}} = 1$ represents complete obscuration by the torus while $f_{\rm{obscured}} = 0$ represents no obscuration.

Fig.~\ref{FigObscurationByTorus} represents the obscuration factor of the accretion disc for $g$ and $y$ bands.
A vertical dashed bar represents the inclination angle coincident with the half-opening angle of the torus (e.g. when inclination = 90$\degree -$ half-opening angle).
Since the flux density of the accretion disc is relatively centralized, obscuration typically occurs at inclinations greater than this value.
The majority of the accretion disc's radiation comes from a region $\mathcal{O} (100) r_{\rm{g}}$, leading to the smooth onset of obscuration after $i = 60 \degree$.
The exact onset is model dependent.
Once the torus completely obscures the accretion disc, the colour difference is due to the wavelength dependence of the extinction profile and the column density of the dust along a sight line to each region of the disc. 

The lower panel of Fig.~\ref{FigObscurationByTorus} illustrates the simulated colour difference $y - g$.
Prior to the torus obscuring the accretion disc, there is no colour dependence with inclination for the accretion disc's radiation.
Once the torus starts obscuring the disc, the nature of the colour dependence is two-fold.
The first effect comes from the differential obscuration by the torus due to the differing sizes of the optical accretion disc.
The second effect comes from the wavelength dependence of the dust's absorption.
In this model, the obscuring torus may be responsible for a colour change of up to a magnitude, though realistic torus densities and absorption profiles will differ.

\section{Correlated Variability Modelling}
\label{VariabilitySection}

AGN typically exhibit correlated variability across their optical and UV light curves. 
A simple way to explain this is the lamp-post model, where a compact X-ray emitter (e.g. the corona) near the SMBH releases high energy photons due to inverse Compton scattering of seed photons from the inner region of the accretion disc~\citep{Collin03, Cackett07, Jha22, Yuk23}.
The driving variability is then associated with variations in the accretion rate at the innermost region of the accretion disc, which is assumed be more sensitive than variations on the larger scales.
The driving source from this point is detailed in Section~\ref{DrivingSignalsection}.
The accretion disc's response to the driving signal is described in Section~\ref{TFSection} and the time lag spectrum is increased by the diffuse continuum as described in Section~\ref{diffuse_continuum_section}.
This variable accretion disc is then assumed to drive the BLR as described in Section~\ref{BLRTransferFunctionSection}.
Combining each of these builds up the time variable source across optical wavelengths in \texttt{Amoeba}.

\subsection{Driving Signal}
\label{DrivingSignalsection}

AGN variability has been traditionally modelled as a stochastic signal, such as a Damped Random Walk (DRW)~\citep{Kelly09} or a Damped Harmonic Oscillator (DHO) ~\citep{Kelly14, Kasliwal17, Moreno19}.
This is done in order to understand relationships between simple model parameters and observations.
These are Continuous-time AutoRegressive Moving Average (CARMA) processes~\citep{Brockwell02, Koen05, Kelly14}. 
The DRW is the simplest CARMA process parameterized by only two values: the characteristic time scale $\tau_{\rm{DRW}}$ and the long term variability described by the asymptotic structure function, SF$_{\infty}$. 
It has a Power Spectral Density (PSD) $\propto f^{-2}$ at high frequencies~\citep{Bauer09} and becomes flat at lower frequencies~\citep{macleod10}.
The DHO is a second-order CARMA process parameterized by four values, leading to greater flexibility of the PSD~\citep{Kelly14}, and has been used to provide better statistical fits and also finds tighter correlations between model parameters~\citep{Kasliwal17, Moreno19}.

It is worth noting that optical variability may be modelled two different ways. 
Optical light curves may be directly fitted by CARMA processes to try to correlate parameters.
This is a statistical approach, but it has been shown that more parameters are required than the DRW case to model variability~\citep{Kasliwal15}.
Codes have also been developed in order to understand the correlation between light curves directly.
\texttt{JAVELIN}~\citep{Zu11} fits a DRW to the bluest light curve and assumes that each other light curve may be represented by a kernel defined as a lagged and broadened top hat function.
Alternatively, forward modelling approaches such as the lamp-post model can generate a natural correlation between wavelengths by allowing the accretion disc to reprocess an X-ray driving signal~\citep{Collin03}. 
This introduces a transfer function between the driving and reprocessed signal which encodes the physics and geometry of the system.
The driving variability is then convolved with the transfer function~\citep{Cackett07, Starkey16, Kammoun19}.
\texttt{CREAM}~\citep{Starkey16} takes advantage of this by reconstructing the driving variability as a Fourier series and then uses thin disc transfer functions.
The convolution with a transfer function affects the power spectrum of the signal, so fitting a CARMA process to a UV/optical light curve may capture different information from fitting a bending power law to the X-ray signal.~\citep{Uttley02, Papadakis04, Paolillo17, Paolillo23}. 

Any PSD may be converted into a stochastic signal by following the method of~\citet{Timmer95}.
\texttt{Amoeba} uses the method defined in section 3 of ~\citet{Timmer95} in order to generate a driving signal from any power spectrum.
By using a bending power law for a model of the coronal emission and a transfer function calculated through the lamp-post model, we build a physically motivated connection between the X-ray driving signal and optical variability~\citep{McHardy04, Uttley05, Oneill05, Markowitz10, Yang22, Yuk23}.

\subsection{Accretion Disc Transfer Function}
\label{TFSection}

Intrinsic variability of an AGN accretion disc may arise from various sources.
We focus on allowing a compact corona to act as the driving source of variability.
However, we note that there are other potential sources of variability, such as time-variable accretion rates at large scales which may lead to long negative lags~\citep[e.g.][]{Secunda23}.
These are not implemented at this time and are planned as the subject of a future update.

The time lag $\tau_{\rm{x}}$ between a point source and all positions on the accretion disc may be calculated in relative cylindrical coordinates.
This is defined as the intersection of isodelay surfaces with positions on the accretion disc as in ~\citet{Sergeev05}, where we modify it to use relative coordinates with respect to the source:
\begin{equation}
\label{timelageq}
\begin{split}
    c \tau_{\rm{x}} (R', \phi', Z' | i) = \sqrt{Z'^{2} + R'^{2}} - Z' \cos(i) - R' \sin(i) \cos(\phi') 
\end{split} 
\end{equation}
where primes denote relative cylindrical coordinates with respect to a source.
Note the sign changes from equation 2 in \citet{Sergeev05}, which are due to the transformation to our defined coordinate system.
For an inclined accretion disc, the azimuth direction closest to the observer (defined to be $\phi = 0\degree$) experiences the shortest time lag, while the azimuth pointing away from the observer ($\phi = 180\degree$) experiences the longest lags.
Furthermore, coordinates above the source experience shorter time lags when compared to those below the source.
When a disc is viewed face on, there is no dependence on the $\phi$ coordinate since the parabolic isodelay surfaces intersect the accretion disc's mid-plane as a series of concentric circles.
We note that Equation (\ref{timelageq}) is applicable to off-axis sources as well after the same change to relative coordinates around the source.

The accretion disc's time dependent temperature profile may be computed by allowing $L_{\rm{x}} \rightarrow L_{\rm{x}}(t)$ and considering the time lags $\tau_{\rm{x}}$:
\begin{equation}
\begin{split}
  \label{timedependenttemp}
  T^{4}_{\rm{obs}}&(R, \phi, Z, t | i, H_{*}) =   T_{\rm{D+W}}^{4}(R) \\ + & \int_{-\infty}^{\infty} dt \left( \frac{(1-A_{*}) L_{x}(t)}{4\pi \sigma_{\rm{sb}} r^{2}_{*}}\right) \cos(\theta_{\rm{x}}) \delta(t-\tau_{\rm{x}}(R, \phi, Z - H_{*} | i)) \\
\end{split}
\end{equation}
where $R, \phi$ is the coordinate of the disc, $i$ is the inclination of the disc, $\delta$ is the Dirac delta function, $t$ is the time, $H_{*}$ is the height of the lamp-post, $r_{*}$ is the distance to the source, and $\theta_{\rm{x}}$ is the angle of incidence of the irradiating photons on the accretion disc.
In the case of a flat accretion disc and a lamp-post located on the axis of rotation, we may define $r_{*}^{2} = R^{2} + H_{*}^{2}$ and $\cos(\theta_{\rm{x}}) = H_{*} / r_{*}$.
We note that the temperature here is the effective temperature that would be observed while including all time lags relative to the observer; not the temperature of the disc within its own frame of reference.

The response of the accretion disc to some instantaneous flare of the corona defines the transfer function.
This is the kernel between any time variable source at the lamp-post's position and the accretion disc's response.
The transfer function for a flat accretion disc with respect to a driving signal by a source above the SMBH may be defined as:
\begin{equation}
\begin{split}
    \label{TransferFunction}
    \Psi(\tau_{\rm{x}}, \lambda &| i, H_{*}) = \\
    &\iint\limits_{s} ds \int_{-\infty}^{\infty} d\tau' \frac{\partial B(T, \lambda)}{\partial T} \frac{\partial T}{\partial L_{\rm{x}}} \delta(\tau' - \tau_{\rm{x}}(R, \phi, -H_{*} | i)) ,
\end{split}
\end{equation}
where the surface integral is carried out over the accretion disc and the integral over $d\tau'$ is carried out for all time lags ~\citep{Cackett07}.
We note that we allow the integration over all times to account for future implementation of long negative lags.
Changing the lower bound from zero does not affect the calculation of traditional time lags due to the Dirac delta function. 
The radial bounds of integration are defined as $r_{\rm{in}}$ and $r_{\rm{out}}$. 
The value of $r_{\rm{in}}$ may be taken to be $r_{\rm{ISCO}}$ for the thin disc model, further out if the truncated accretion disc model is used~\citep[e.g.][]{Serafinelli23}, or zero if another temperature model is used (e.g. a Gaussian profile). 
$r_{\rm{out}}$ is a model parameter, typically of order 10$^{3}$ to 10$^{4}$ $r_{\rm{g}}$ for SMBH associated with AGN.
Alternatively, if an optically thick torus is modelled, $r_{\rm{out}}$ can be taken to be the point when the sublimation temperature $T_{\rm{sub}}$ is reached (e.g. Fig.~\ref{Fig_sublimation_radius}).

\begin{figure}
    \centering
    \includegraphics[width=0.47\textwidth]{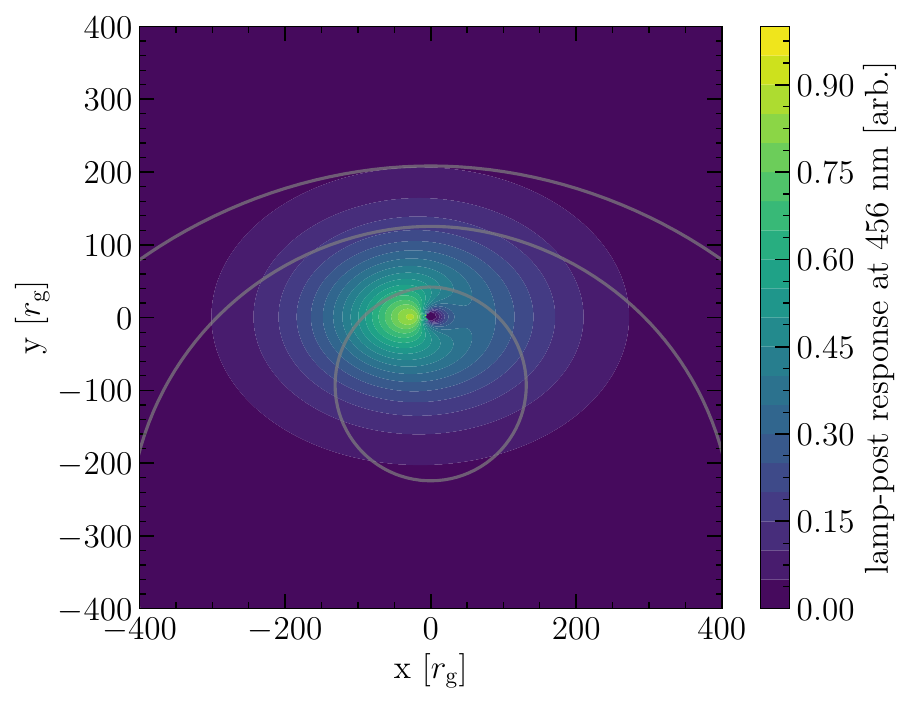}
    \caption{Accretion disc response projected into the source plane for $i = 45\degree, H_{*} = 10 r_{\rm{g}}$ and redshift 0. Isodelay contours are placed at 100, 300, and 500 $r_{\rm{g}}$. We note the response distribution is more widespread than the surface flux distribution of Fig.~\ref{FigSurfaceBrightness}.}
    \label{IMGisodelaysurface}
\end{figure}

We define the response of the disc as the product of the partial derivatives in Equation (\ref{TransferFunction}).
The response is subject to the same effects that are applied to the accretion disc's emission (wavelength shifting, relativistic beaming, and geometric distortions due to curved geodesics around the central SMBH).
\texttt{Amoeba} constructs the response by calculating each partial derivative of Equation (\ref{TransferFunction}) while considering the adjusted wavelengths, amplifications, and apparent shifts for each pixel on the accretion disc.

Fig.~\ref{IMGisodelaysurface} illustrates the central 400 by 400 $r_{\rm{g}}$ region of the surface response of an accretion disc when inclined at inclination $i = 30\degree$. 
Each pixel represents its relative response due to fluctuations of the point source corona located at a position of 10 $r_{\rm{g}}$ above the SMBH.
Isodelay contours are placed at 100, 300, and 500 $r_{\rm{g}}$ intervals which represent the time lag $\tau_{\rm{x}}$ from the corona--disc--observer path with respect to the direct path between the corona and the observer.
The transfer function $\Psi(\tau_{\rm{x}}, \lambda)$ is constructed by binning pixels based on $\tau_{\rm{x}}$, weighted by the response, and represents the disc's response to an impulse by the corona.
We note that a single transfer function does not explicitly define the time lag between two optical wavelengths, but this may be derived from multiple transfer functions.

\begin{figure}
    \centering
    \includegraphics[width=0.47\textwidth]{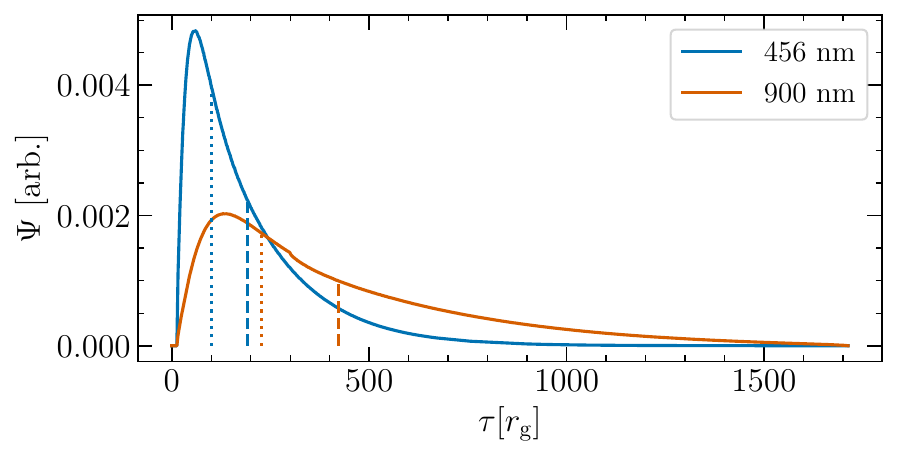}
    \caption{Calculated accretion disc transfer functions for 456 and 900 nm for the same parameters as in Fig.~\ref{IMGisodelaysurface}. The geometric mean time lags are denoted by dashed vertical bars and the centroids of the transfer functions are represented by the dotted vertical bars.}
    \label{IMGdisctransferfunction}
\end{figure}

Fig.~\ref{IMGdisctransferfunction} illustrates the constructed transfer functions from the disc in Fig.~\ref{IMGisodelaysurface}, evaluated at observed wavelengths 456 and 900 nm. 
These transfer functions tend to have a common shape: a steep initial rise followed by a long trailing response.
Each transfer function is normalized such that they integrate to unity (e.g. $\int_{0}^{\infty} d\tau' \Psi(\tau', \lambda) = 1$ for all $\lambda$).
This is imposed so the convolution does not introduce extra power with wavelength dependence.
However, we note that this normalization removes information regarding the relative responses of the accretion disc between different wavelengths~\citep[see][for an alternative approach]{Kammoun19}.

The mean time lag $\bar{\tau}_{\rm{x}}(\lambda)$ is defined as the geometric mean:
\begin{equation}
    \label{mean_time_lag_eq}
    \bar{\tau}_{\rm{x}}(\lambda) = \frac{\int_{0}^{\infty} d\tau' \tau' \Psi(\tau', \lambda)}{\int_{0}^{\infty} d\tau' \Psi(\tau', \lambda)} ,
\end{equation}
and is illustrated in figures by dashed vertical lines.
These are wavelength dependent such that redder wavelengths have longer time lags.
This is similar to surface flux distributions having a wider effective radius at redder wavelengths.
Time lags scale proportionally with the black hole mass, and the skewness of the transfer function scales with inclination.
The skew's impact has been explored within the context of traditional time lag fitting software such as \texttt{JAVELIN}, \texttt{PyCS}, and \texttt{CREAM}, where the importance of accurate transfer functions are emphasized in~\citet{Chan20}.
Equation (\ref{mean_time_lag_eq}) should receive careful consideration to how it is interpreted if long negative lags are considered in the transfer function.

Additionally, the centroid of the transfer function can be considered as this is representative of the centroid value of the cross-correlation function~\citep{Koratkar91}.
We define the centroid as the central time lag of the full width at half-maximum.
To highlight the difference between the geometric mean and centroids of our accretion disc transfer functions, we plot both of them in each figure.
In all cases, dashed lines represent the geometric means while the dotted lines represent the centroids.
This is of particular importance when using transfer functions from inclined discs, since the geometric mean has very little inclination dependence, but the centroid is more significantly affected by inclination.
The geometric mean and the centroid coalesce when the transfer function takes a symmetric form.

\begin{figure}
    \centering
    \includegraphics[width=0.47\textwidth]{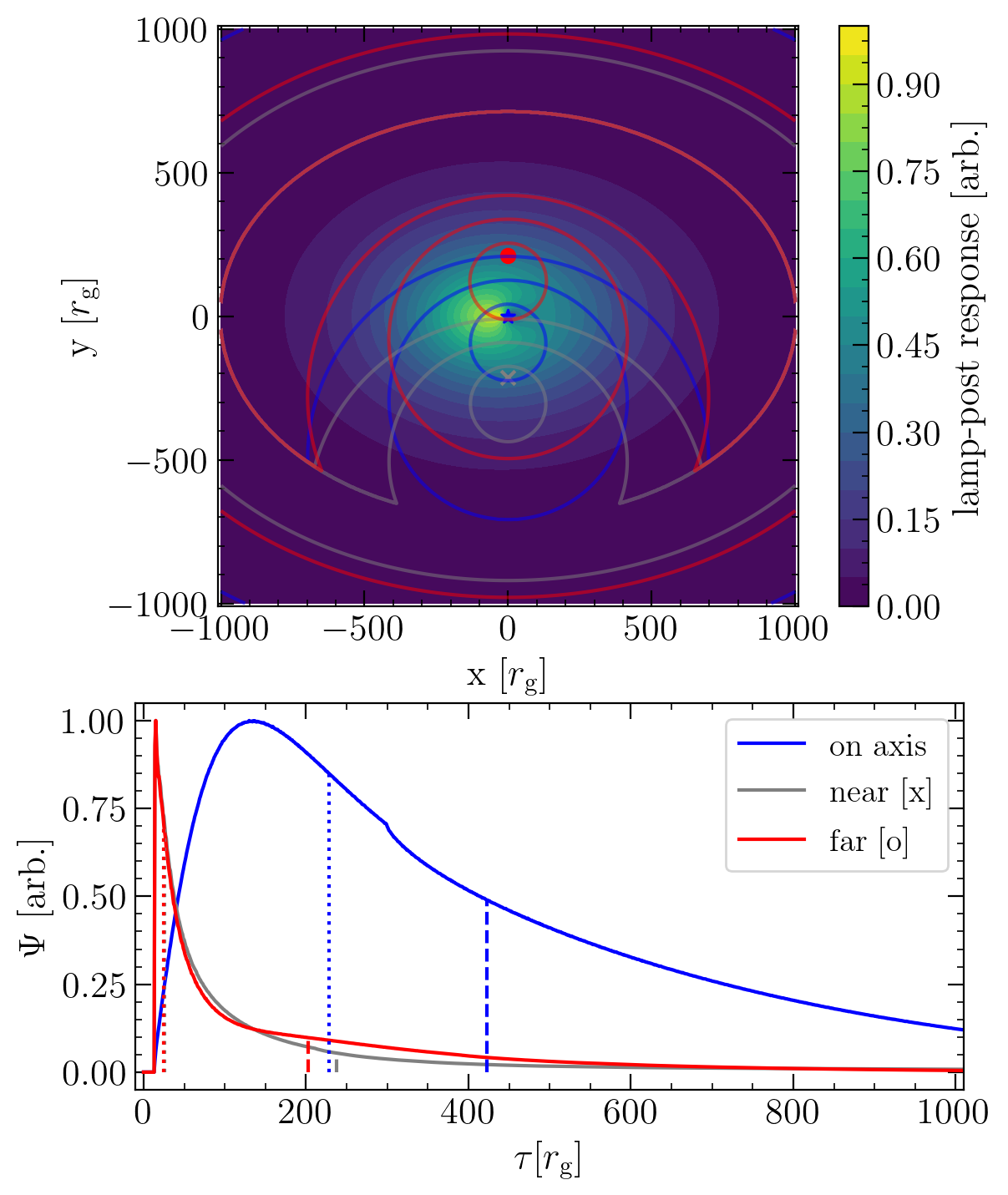}
    \caption{Sample transfer functions for a thin accretion disc are derived from central and off-axis sources at $\lambda = 900$ nm. Two choices for the off-axis sources are located 200 $r_{\rm{g}}$ off the axis of symmetry. The circle (o) denotes the near side closer to the observer, while the cross (x) represents the far side. Isodelay contours from each position are placed at 100, 300, and 500 $r_{\rm{g}}$. We note that the accretion disc's response is only shown for the traditional lamppost source.}
    \label{fig_lamppost_off_axis}
\end{figure}

While a flexible and physically motivated temperature profile is provided in Equation (\ref{fulltempprofile}), \texttt{Amoeba} is flexible enough to produce responses for any effective temperature distribution from any position of the source.
Fig.~\ref{fig_lamppost_off_axis} illustrates accretion disc transfer functions generated for both the usual lamp-post model as well as off-axis sources. 
The upper panel's colour map represents the relative response of the accretion disc to a flare due to a point source located at 10 $r_{\rm{g}}$ above the SMBH, positioned at the blue star.
The responses of the accretion disc to off-axis sources are not plotted, but peak at a relatively localized region nearest to the source as suggested by the shape of their transfer functions.
Isodelay surfaces of 100, 300, and 500 $r_{\rm{g}}$ are plotted for this source in blue.  
Additional isodelay surfaces for fiducial off-axis sources located at 300 $r_{\rm{g}}$ are plotted in grey and red for near and far sources denoted by the cross (x) and circle (o), respectively.
The lower panel represents the relative response functions of a flare located at each position.
The mean time lag in both off-axis cases is less than the traditional lamp-post location labelled as the star due to the increased response of the disc at lower temperatures.
The centroids of both off-axis response functions are significantly smaller than the traditional case.

Within the lamp-post model, the driving variability from a coronal X-ray source is taken to be a point source.
We note that this is an approximation, as treating an extended source imposes multiple computational challenges.
Some of these include:
\begin{itemize}
    \item The emitting regions must be integrated over at each time step.
    \item Radiative transfer through the corona should be considered.
    \item A correlation model between different regions of the corona should be considered.
\end{itemize}
Confronting these challenges may be the topic of a future update.

\subsection{Diffuse Continuum}
\label{diffuse_continuum_section}
It has been shown that the diffuse continuum can significantly affect inter-band time delays.
This is described in~\citet{Korista01} where the diffuse continuum arising from hydrogen was used to explain larger accretion disc sizes when compared to theory.
Photoionization codes such as \texttt{CLOUDY} support this claim by computing the response of clouds with many ionic species.
The diffuse continuum emission and lag spectrum is primarily dominated by the Balmer and Paschen continua~\citep{Ferland17}. 
Other spectral features may be seen to a lesser extent.
Notably, observations show that the Lyman continuum is never observed~\citep{Korista19}.

\texttt{Amoeba} does not attempt to replace codes designed to fit the features of the diffuse continuum.
Instead, it takes the prescription from~\citet{Korista19}: for a given diffuse continuum time delay spectrum $\tau_{\rm{DC}}(\lambda)$ and weighting factor, the diffuse continuum increases the mean time lag by a factor $\bar{\tau}_{\rm{DC}}$~\citep[See also][]{Hernandez20, Hagen24}.
This can simulate the increased time lags associated with the diffuse continuum.

Furthermore, Amoeba can assume some radial distribution of clouds which corresponds to the physical distribution of the emitting region.
In doing so, the flux density from the diffuse continuum may be integrated into the microlensing pipeline to have a fully self-consistent model between all components building up the AGN model (e.g. see Section~\ref{MicrolensingSection}).
We note that this does not currently consider the emission spectra as a function of radius, which may be the target of a future update.

\subsection{BLR Transfer Function}
\label{BLRTransferFunctionSection}

The continuum is assumed to provide the optical variability that drives the BLR response. 
While the full accretion disc will contribute to the fluorescence of the line-emitting species, we approximate the optical light curve's origin as a point source located at the central black hole.
We note that there is considerable room for growth within in this model to account for the extended geometry of the disc within the transfer function, and will be the focus of a future update.
Transfer functions have been calculated for multiple BLR geometries, and may lead to either single or double peaked transfer functions~\citep[e.g. thick clouds, spherical biconical outflows, thin discs][]{Peterson01}.

The time lags between the accretion disc and the BLR are defined as in Equation (\ref{timelageq}).
We approximate the time lag between the accretion disc and the BLR by calculating all time lags with respect to the SMBH, where the average emission from the accretion disc is assumed to originate.
In reality, this will have some dispersion relative to the size of the accretion disc.

The BLR transfer function may now be expressed as:
\begin{equation}
\begin{split}
    \label{BLRTransferFunction}
    \Psi_{\rm{BLR}}(\tau, \lambda | i) = &\iiint\limits_{V} dV \int d\tau' \int d\lambda' \frac{\partial L_{\rm{BLR}}}{\partial L_{\rm{con}}} \\
    & \times \delta(\tau - \tau'(R, \phi, Z | i)) \delta(\lambda - [F_{\rm{line}}(\lambda')]) ,
\end{split}
\end{equation}
where the volume integral is taken over the entirety of the emitting BLR region and we have exchanged the partial derivatives for the (instantaneous) fluorescence of an emission line $L_{\rm{BLR}}$ due to the driving continuum $L_{\rm{con}}$, defined as $\frac{\partial L_{\rm{BLR}}}{\partial L_{\rm{con}}}$.
The Dirac delta function $\delta(\lambda - [F_{\rm{line}}(\lambda')])$ picks out only the emission from the line profile [$F_{\rm{line}}(\lambda')$] which match the simulated wavelength $\lambda$.
The line profile is assumed to only depend on the line-of-sight velocity with respect to the observer.

\begin{figure}
    \centering
    \includegraphics[width=0.47\textwidth]{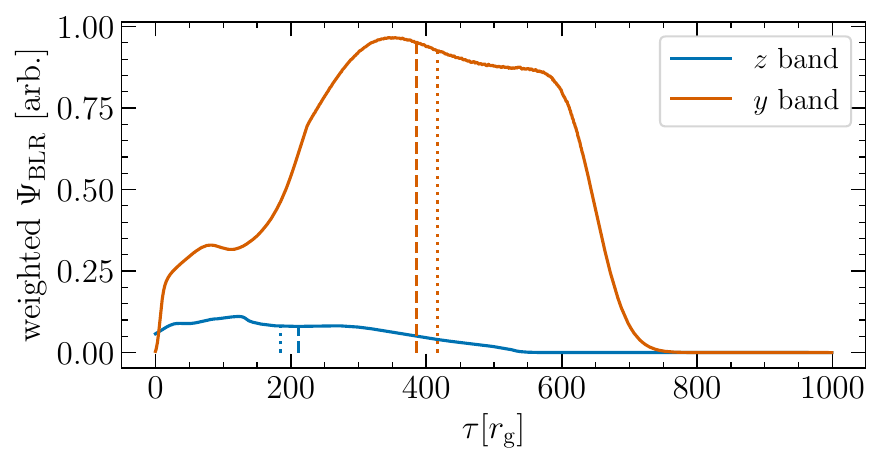}
    \caption{Response function of the velocity selected BLR with respect to a time varying continuum as modelled in Section~\ref{BLRsection}. Mean time delays are shown as dashed lines, centroids are denoted with dotted lines, and each function is weighted by its relative response.}
    \label{BLRTF}
\end{figure}

Fig.~\ref{BLRTF} illustrates the response of our model emission line, where we compare the response function as seen by the LSST $z$ and $y$ bands.
These corresponding response functions are representative of the densities in Fig.~\ref{ProjectedBLR} and weighted by the amount of material contributing to each filter.
Only some regions with relatively high line-of-sight velocity are projected into the $z$ band, leading to an asymmetric projection.
This asymmetry propagates into a different response function of the BLR in this band.
Depending on which velocities are required to enable contamination in a simulated band, the contamination can significantly change.

\begin{figure}
    \centering
    \includegraphics[width=0.47\textwidth]{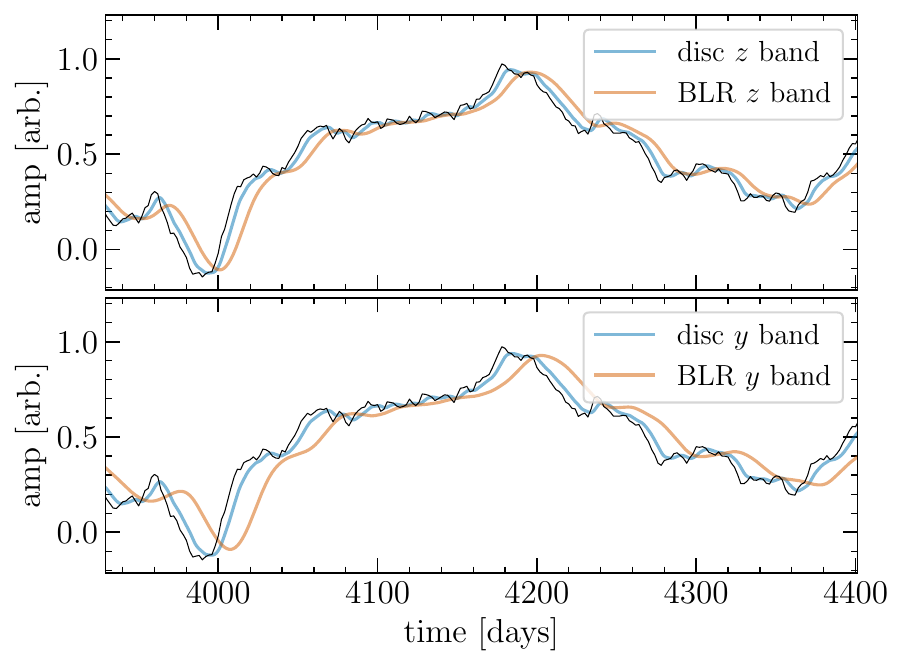}
    \caption{Modelling of the propagation of an intrinsic signal through the accretion disc and BLR for the $y$ and $z$ filters. This reprocessed signal drives the BLR in each band as described by the BLR transfer functions in Fig.~\ref{BLRTF}. The mass of the SMBH was increased to $10^{8.5} M_{\odot}$ to accentuate the lags between the driving source represented by the thin black line, the disc, and the BLR.}
    \label{IMGSignaldiscBLR}
\end{figure}

Fig.~\ref{IMGSignaldiscBLR} illustrates a stochastic signal driving the response of multiple components of the AGN model.
The corona is represented by the X-ray light curve, originates at 10 $r_{\rm{g}}$ above the SMBH, and is considered compact.
The accretion disc's signal in the $z$ and $y$ bands are calculated and allowed to drive the BLR in each case.
We note that the relative emission from the BLR in the $z$ band is $\sim$ 11.5 per cent that of the $y$ band due to the relatively small region which contributes.
This must be considered when combining light curve components to create a self-consistent light curve.

\begin{figure}
    \centering
    \includegraphics[width=0.95\linewidth]{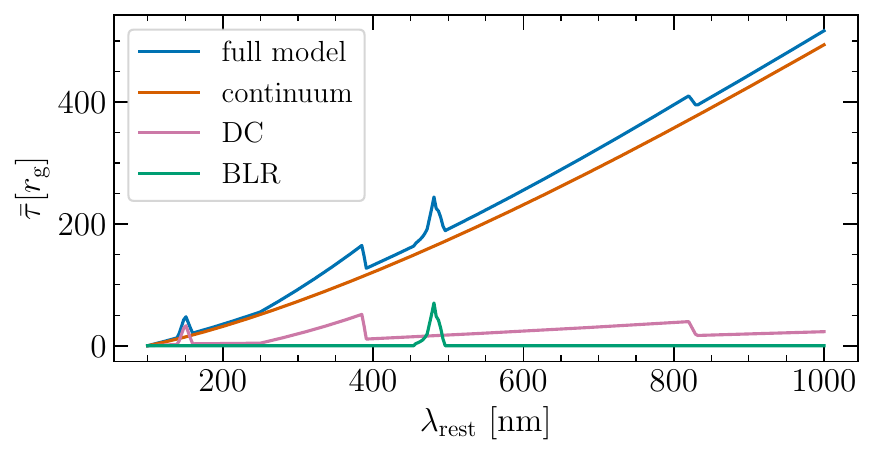}
    \caption{Example time lag spectrum which includes the geometric mean of the optical continuum, diffuse continuum, and a broad line region emission line. The diffuse continuum is taken to have a minimum radius of 2 light days, a constant cloud density profile, emissivities resembling those of NGC 5548 with scale factor 0.5~\citep{Korista19}. Other parameters include $M_{\rm{BH}} = 10^{8.4} M_{\odot}, ~i = 45\degree, ~z_{\rm{s}} = 0.1, v_{\infty} = 0.2$.}
    \label{fig_diffuse_continuum_spectrum}
\end{figure}

Joining together the driving signal, accretion disc, diffuse continuum, and broad line region leads to a model which can simulate time lags across the optical spectrum.
We note that while the lamp-post model assumes that the driving signal comes from the stochastic X-ray corona, the observed X-ray flux rarely corresponds to the expected driving signal. 
Therefore, the observable is the inter-band time lags within the light curves.

Fig.~\ref{fig_diffuse_continuum_spectrum} illustrates a mock time lag spectrum with respect to the 100 nm response for all components of the model including the accretion disc's continuum, the diffuse continuum contribution, and the BLR.
Each component is plotted along with the total spectrum and weighted by its relative contribution to the increase in mean time lag.
We note that mean time lags are computed as the geometric mean over the associated transfer function for the accretion disc and BLR, and as the mean increase in lag for the diffuse continuum.
It is evident that both the diffuse continuum and BLR have the ability to substantially increase the inter-band mean time lag at certain wavelengths.
We note that while these spectral features are highly model dependent, \texttt{Amoeba} aims to include most models in its framework.

\section{Extrinsic Variability due to Microlensing}
\label{MicrolensingSection}

Strongly lensed objects have the potential to exhibit microlensing due to compact objects within the lensing galaxy~\citep[See][and references therein]{Vernardos23}.
These objects within the lensing galaxy (e.g. stars) may gravitationally influence photons on their path to the observer. 
The impact microlensing has on an image may be represented by a magnification map. 
This map encodes the total magnification from randomly distributed microlenses in the source plane.
These microlensing maps may be parametrized by a few key values related to the lensing potential at the image's position: Convergence, shear, and smooth matter fraction. 

The important size scale for microlensing is the Einstein radius of the microlenses ($\theta_{\rm{E}}$), which is defined in radians as:
\begin{equation}
    \theta_{\rm{E}} = \sqrt{\frac{4GM_{\rm{ml}}}{c^{2}}\frac{D_{\rm{ls}}}{D_{\rm{ol}}D_{\rm{os}}}}
\end{equation}
where $M_{\rm{ml}}$ is the average microlens mass, and the distances $D_{\rm{ol}}, D_{\rm{os}}, \text{ and } D_{\rm{ls}}$ are the angular diameter distances of the observer-to-lens, observer-to-source, and lens-to-source respectively. 
These distances depend on the redshifts of the lens and source, and the assumed cosmology.
It is convenient to define the Einstein radius in the source plane as $R_{\rm{E}} = D_{\rm{os}} \theta_{\rm{E}}$.

Once the distance scales are determined, the magnification map represents the spatial distribution of magnification due to microlenses in the source plane.
This magnification is achromatic in the sense that it affects each wavelength equally, and the total microlensed flux may be calculated by the convolution between the magnification and flux distributions:
\begin{equation}
    \label{microlensing_magnification}
    F_{\rm{ml}}(\lambda) = \mu_{\rm{ml}} * F(\lambda) 
\end{equation}
where $\mu_{\rm{ml}}$ is the magnification due to microlenses and $F(\lambda)$ is the flux distribution of the accretion disc at wavelength $\lambda$.

\texttt{Amoeba} will simulate microlensing between surface flux distributions of AGN and externally provided magnification maps.
Magnification maps may be constructed using methods such as Inverse Ray Tracing~\citep{Kayser86}, Inverse Polygon Mapping~\citep[][]{Mediavilla06}\citep[see also the Fast Multipole Method extension][]{Jimenez22}, GPU extensions of these methods \footnote{\href{https://github.com/weisluke/microlensing}{https://github.com/weisluke/microlensing}} (Weisenbach et al. in prep.), or other techniques. 
Representative microlensing maps may be obtained from the GERLUMPH~\citep{Vernardos14} database, which were generated using \texttt{GPU-D}~\citep{Thompson10} over an extensive parameter range of convergence, shear, and smooth matter fraction. 

\begin{figure}
    \centering
    \includegraphics[width=0.47\textwidth]{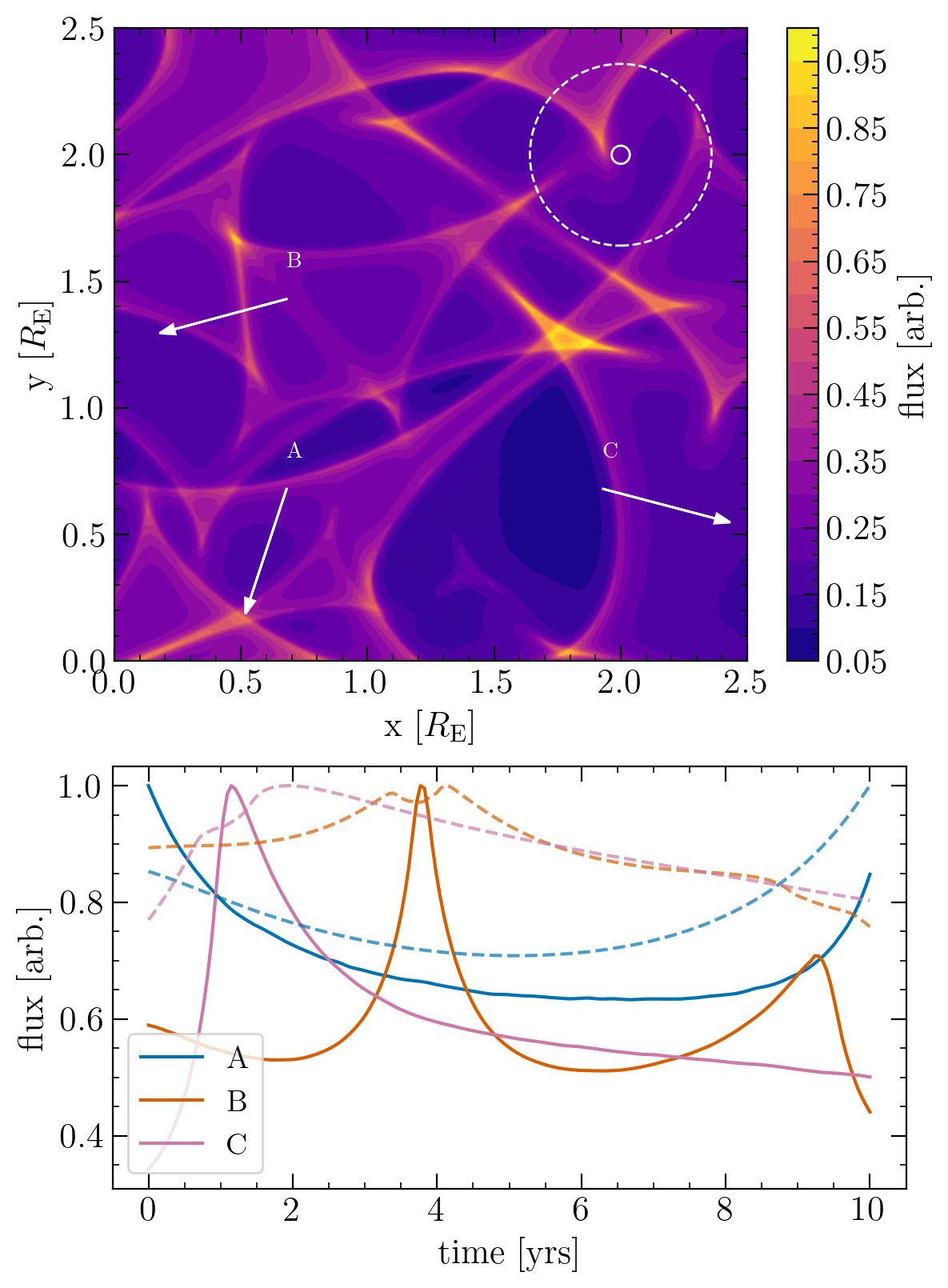}
    \caption{Top panel: Convolution between a magnification map and surface flux distribution for $M_{\rm{BH}} = 10^{9} M_{\odot}$ observed at $\lambda_{\rm{obs}} = 900$ nm. Solid and dashed circles with radii of 100 $r_{\rm{g}}$ are drawn to represent the relative size scales of sources with $M_{\rm{BH}} = 10^{8} \text{ and } 10^{9} M_{\odot}$, respectively. Other parameters include $z_{\rm{l}} = 0.5$, $z_{\rm{s}}$ = 1.0, $M_{\rm{ml}} = M_{\odot}$, convergence = 0.30, shear = 0.39, smooth matter fraction = 0.0, $v_{\rm{e}}$ = 600 km s$^{-1}$. Sample light curves are extracted along the arrows. Bottom panel: Normalized light curve across each track for two surface flux distributions, where solid lines represent the light curve for $M_{\rm{BH}} = 10^{8} M_{\odot}$ and dashed lines represent $M_{\rm{BH}} = 10^{9} M_{\odot}$. Microlensing is more pronounced for smaller accretion disc sizes, and the ISCO feature is apparent in the larger disc's light curve.}
    \label{FigConvolutionWithTracksAndLightCurves}
\end{figure}

To simulate microlensing, \texttt{Amoeba} performs the following steps for any magnification map:
\begin{enumerate}
    \item Calculate $R_{\rm{E}}$ for a given cosmology to match the pixel size scales of the source and microlensing map using \texttt{Astropy}~\citep{astropy13, astropy18, astropy22}. 
    \item The accretion disc image is scaled to match pixel sizes between the simulated accretion disc and the magnification map using \texttt{SciKit-Image}~\citep{VanderWalt14}.
    \item The convolution between the surface flux distribution is calculated using the Fast Fourier Transform (FFT) package from \texttt{SciPy}~\citep{Scipy20}.
    \item The SMBH location is used as the point of reference so all components of the AGN are convolved with respect to the center.
    \item For a given effective velocity and duration, light curves are extracted from the convolution.
\end{enumerate}
The SMBH is used as the reference point in all microlensing simulations in order to consistently microlens each component.
This is important when comparing different accretion discs (e.g. Fig.~\ref{FigConvolutionWithTracksAndLightCurves}) or when including contributions from a BLR model (e.g. Fig.~\ref{FigdiscBLRMicrolensed}).
Care must be taken to avoid convolution artifacts near the boundaries.
To avoid edge artifacts, \texttt{Amoeba} will only draw random paths from the central region.
We note that if the source is too large, a large fraction of the convolution will be excluded.
However, for such sources, microlensing will have little to no impact and the magnification will converge to the macro-magnification of the image.

Within \texttt{Amoeba}, a light curve from the convolution is defined by three values: an origin point, a direction, and a total distance to travel.
The distance travelled is defined by the effective transverse velocity in the source plane $v_{\rm{eff}}$.
Typical velocity distributions for realistic systems are found to be $\mathcal{O} (100)$ km s$^{-1}$.
For extraordinary systems such as the Einstein Cross (Q2237+0305), the effective velocity may be as high as a few thousand km s$^{-1}$~\citep{Huchra85, Kayser86, Wambsganss90, Kochanek04, Eigenbrod08, Mediavilla15a, Mediavilla15b, Neira20}.

Fig.~\ref{FigConvolutionWithTracksAndLightCurves} illustrates the convolution between a magnification map and a surface flux distribution of an accretion disc with $M_{\rm{BH}} = 10^{9} M_{\odot}$.
A second convolution was performed between the original magnification map and a surface flux distribution of an accretion disc with $M_{\rm{BH}} = 10^{8} M_{\odot}$ to compare light curves.
We zoom in on a small section of the 25 x 25 $R_{\rm{E}}$ convolution for illustrative purposes.
Light curves are taken along the arrows labelled A, B, and C on both convolutions.
Each light curve is individually normalized to emphasize the impact of the source size, and is presented in the lower panel.
Regions of high magnification occur when the accretion disc crosses over a caustic fold or cusp: regions that lead to the creation of new micro-images.
This leads to the encoding of the accretion disc's structure in the light curves as described in~\citet{Vernardos_2019}.
Depending on the system configuration, the spatial resolution gained through microlensing has the ability to encode the shadow of the black hole as described in~\citet{Best24} for an appropriate ratio of $r_{\rm{g}} / R_{\rm{E}}$.

\section{Joint Modelling of Intrinsic and Extrinsic Variability}
\label{FullVariabilitySection}

Variability is often used to constrain size scales of AGN components.
Some studies use the AGN's intrinsic variability to determine sizes of the BLR or the accretion disc~\citep[e.g.][]{Kaspi01, Zu11, Paic22}.
Other studies use microlensing in order to probe size scales of the AGN with respect to the lensing galaxy~\citep[e.g.][]{Anguita08, Bate08, Jimenez12}.
In each case, studies treat the other variable component as noise and aim to remove it.
However, there is another component of the variability which may arise from the interplay between the intrinsic variability and microlensing.
This manifests as the microlensing time delay and can affect the reverberation properties of the source~\citep{Tie18, Chan21}.
Furthermore, when multiple components of the AGN are modelled together, the BLR may contaminate the signal which is assumed to only come from the accretion disc~\citep{Netzer22}.
In this section, we will show how consistent modelling of multiple components can impact expectations.

\subsection{Intrinsic Light Curves}

\texttt{Amoeba} is an environment which has the ability to join together many AGN model components, including the reverberating accretion disc, the diffuse continuum, and the broad line region.
It is not restricted as a tool for strongly lensed AGN, which only account for a small fraction of all AGN.

In order to highlight how the diffuse continuum and BLR can contaminate optical bands in future wide-field surveys, we prepare our model AGN with the following parameters: $M_{\rm{BH}} = 10^{8.4} M_{\odot}$, $z_{\rm{s}} = 0.8$, $i = 30\degree$, $R_{\rm{in, BLR}} = 800 R_{\rm{g}}$, $R_{\rm{LOC}} = 1200 \pm 200 R_{\rm{g}}$, and a Novikov-Thorne profile accreting at an Eddington ratio of 5 per cent.
The intrinsic driving signal is taken to be a power law for simplicity, with spectral index of -2. 
We note that this is equivalent to a DRW with a characteristic time scale longer than the 300 day light curve generated.
The BLR is assumed to manifest as a discrete emission line located at 486 nm, which contributes 20 per cent of the continuum emission in the $z$ filter.
The diffuse continuum is modelled with the spectrum of NGC 5548~\citep{Korista19, Hernandez20}, is assumed to be co-spatial with the BLR, and contributes a maximum of 30 per cent of the continuum.
We generate optical signals by using \texttt{Speclite} to model LSST filters using the "LSST2023-*" set of throughputs.

\begin{figure}
    \centering
    \includegraphics[width=0.95\linewidth]{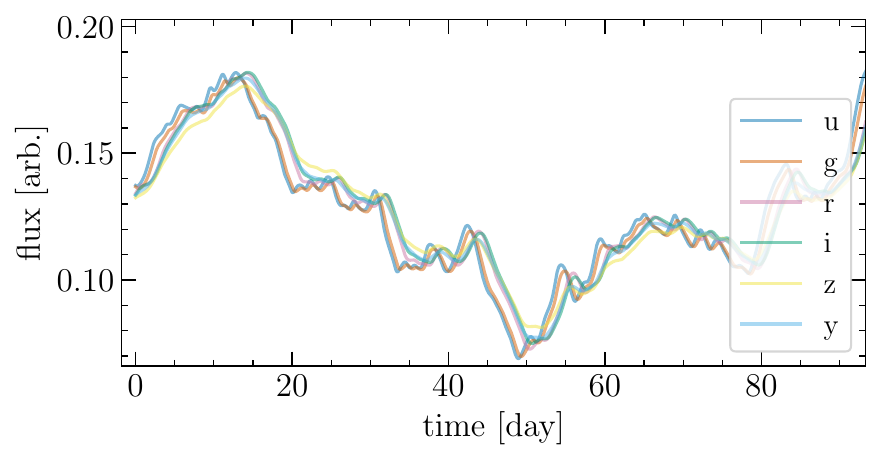}
    \caption{A zoomed set of dense optical light curves generated by propagating a driving signal through each AGN model component. The full light curves last 300 days. All bands are smoothed by the accretion disc transfer functions as described in Section~\ref{TFSection}, shifted by an additional time lag as defined in Section~\ref{diffuse_continuum_section}, then used to drive the BLR signal as described in Section~\ref{BLRTransferFunctionSection}. When the BLR falls in a \texttt{speclite} filter, it is assumed to contribute up to 20 per cent of the continuum emission. This only occurs in the $z$ filter. The diffuse continuum and broad line region increase the complexity in the inter-band time delays.}
    \label{fig_sample_zoomed_light_curve}
\end{figure}

\begin{figure}
    \centering
    \includegraphics[width=0.95\linewidth]{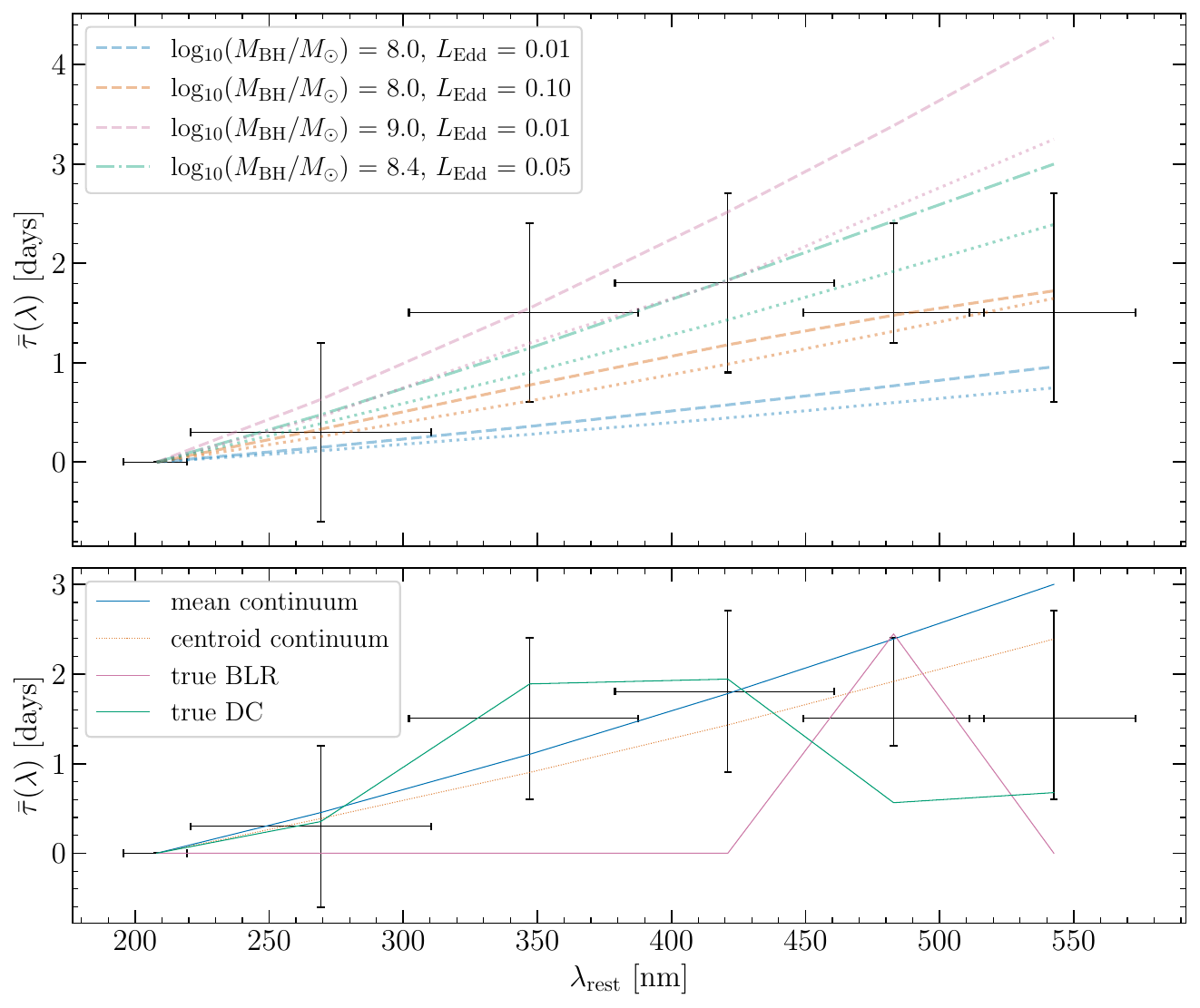}
    \caption{Time lag spectrum for the light curves of Fig.~\ref{fig_sample_zoomed_light_curve} with respect to the $u$ band. Time lags are computed using an implementation of the discrete cross-correlation function (DCF). The top panel has typical model spectrum of time lags for a reverberating thin disc for various combinations of parameters. Dashed lines and dotted lines represent time lags expected from the geometric means and centroids of transfer functions, respectively. The lower panel illustrates the lag spectrum with the true time lag components, scaled by their relative contributions. Here it is evident that the measured time lags using the DCF method are underestimating the expected time lags.}
    \label{fig_time_lag_spectra}
\end{figure}

We illustrate these mock light curves in Fig.~\ref{fig_sample_zoomed_light_curve}, where an intrinsic signal is propagated through each model component in order to produce a set of dense light curves.
We sample them at daily cadence in each band, to simulate an upper limit to intensive monitoring for optical reverberation mapping.
We zoom in on the light curves to highlight some fine details of the light curves, where the diffuse continuum and BLR may significantly affect the time lags between bands.
We note that the modelled BLR emission line falls into the $z$ band, while the diffuse continuum mainly impacts the $r$ and $i$ bands. 
This makes the $z$ band light curve appear more smoothed than the $y$ band, even though it has a shorter rest wavelength.

Fig.~\ref{fig_time_lag_spectra} represents a brief discrete cross-correlation analysis of the light curves from Fig.~\ref{fig_sample_zoomed_light_curve} using the package \texttt{PyDCF}~\citep{Edelson88, Robertson15}.
The measured time delays with respect to the $u$ band are represented as black points, which have error bars corresponding to the filter ranges and the time delay uncertainties.
The top panel illustrates the time delay spectra associated with multiple thin disc models in dashed lines. 
The green dot-dashed line represents the expected time lags associated with the actual simulation parameters.
Expected time lag contributions from the continuum, diffuse continuum, and BLR are represented by coloured curves in the lower panel, which were calculated from the geometric means of the transfer functions.
We note that cross-correlation methods are known to underestimate the time lag with respect to the geometric means of transfer functions~\citep{Chan20} and are instead more representative of the centroids of the transfer functions.
It is evident that the accretion disc model alone does not have the ability to capture all components, and a best fit single component model may over estimate the mass or accretion rate of the AGN.
Furthermore, while the broad emission line signal can be isolated using spectral decomposition techniques, it has only recently become possible to isolate the diffuse continuum~\citep{Korista19, Hernandez20, Nunez23b, Hagen24, Netzer24}.

\subsection{Microlensing Time Delay}
\label{SlowMicrolensingSubsection}

The variability time scales of AGN depend primarily on the sizes of each component.
X-ray variability is known to exhibit rapid fluctuations over very short time scales, while low-ionization emission lines within the BLR vary over much longer time scales.
Within microlensing, the typical time scale is known as the source crossing time which depends on $v_{\rm{eff}}$ and the source size.
Typically, intrinsic variability is much more rapid than microlensing.
We can illustrate the impact microlensing has on the intrinsic variability by assuming that microlensing is not evolving with time.
We note that treating microlensing as stationary is not equivalent to ignoring it.
Constant microlensing will modulate the surface response distribution and the transfer function.
This is especially important for sources near caustics, as the caustic may significantly change the reverberation mapping mean time lags inferred.

The response distributions will be affected by microlensing similarly to the surface flux distributions.
Following Equation (\ref{microlensing_magnification}), the magnified response of the response distribution is defined to be the product ($\mu_{\rm{ml}} \frac{\partial B}{\partial T} \frac{\partial T}{\partial L_{\rm{x}}}$).
We then construct the transfer function as in Equation (\ref{TransferFunction}) with the microlensed response distribution.
Doing so produces what we call a microlensed transfer function.
Mean time lags derived from the microlensed transfer function may then differ with respect to the traditional case.
We note that these microlensed transfer functions will now depend on the microlensing configuration and parameters, in addition to usual geometric and source parameters.

\begin{figure}
    \centering
    \includegraphics[width=0.47\textwidth]{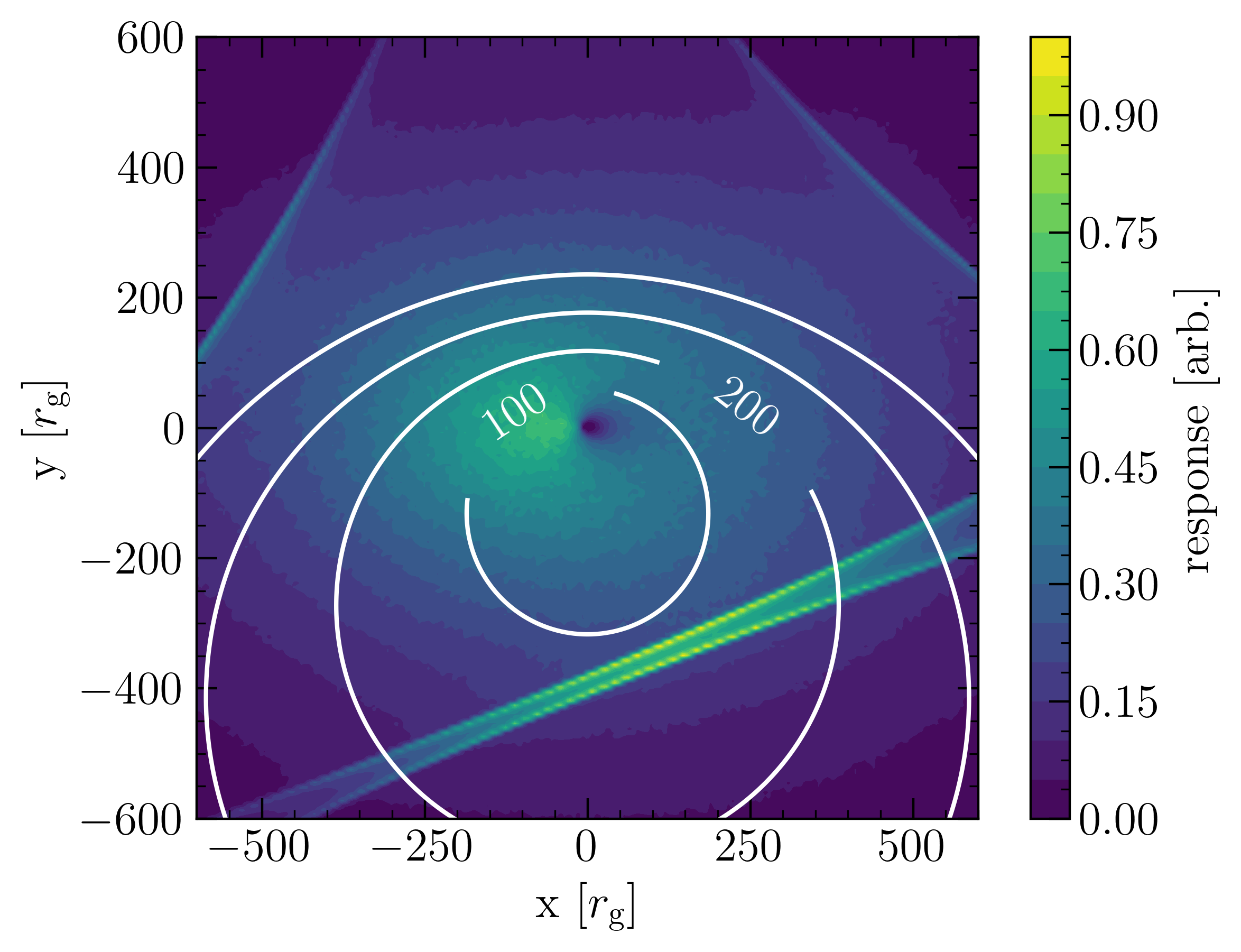}
    \caption{The response of the accretion disc when placed near a microlensing caustic fold for $M_{\rm{BH}} = 10^{8} M_{\odot}$, $i = 30\degree$, $\lambda_{\rm{obs}} = 900$ nm. Other parameters are identical to those described in Fig.~\ref{FigConvolutionWithTracksAndLightCurves}. Representative isodelay contours are placed at time lags of 100, 200, 300, and 400 $r_{\rm{g}}$.}
    \label{figMicrolensedResponseMap}
\end{figure}

\begin{figure}
    \centering
    \includegraphics[width=0.47\textwidth]{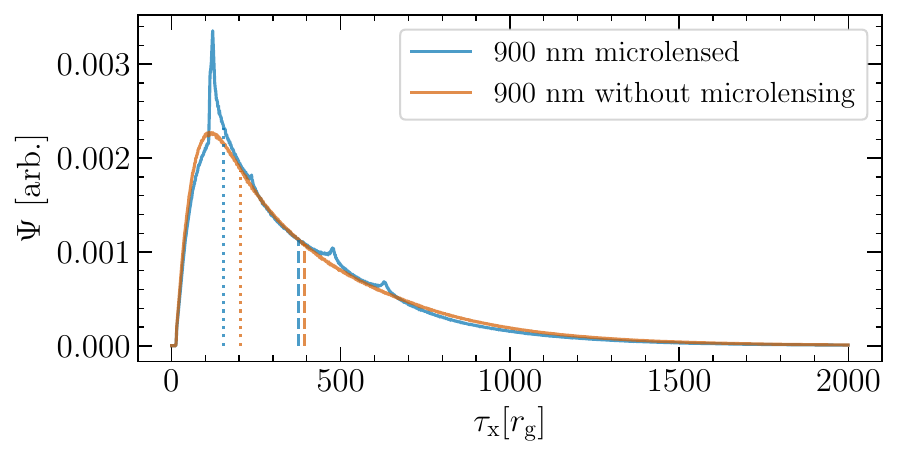}
    \caption{The derived transfer function for the disc illustrated in  Fig.~\ref{figMicrolensedResponseMap}. The geometric means of the transfer functions are presented as dashed lines, while the centroids are represented by the dotted lines.}
    \label{figMicrolenseddiscTF}
\end{figure}

Fig.~\ref{figMicrolensedResponseMap} illustrates the microlensed response of an accretion disc near a caustic fold.
Isodelay contours are projected onto the magnified surface response as in Fig.~\ref{IMGisodelaysurface}.
A caustic fold amplifies regions of the surface response with time lag between 0.5 and 2 days, changing its response distribution. 

Fig.~\ref{figMicrolenseddiscTF} represents the microlensed transfer function from the response described by Fig.~\ref{figMicrolensedResponseMap}.
The magnification due to the caustic fold affects the transfer function and shortens the geometric mean and centroid of the time lag.
Although microlensing itself is achromatic, it differentially magnifies spatially distinct regions of the source.
It is evident that the centroid may be more sensitive to the microlensing time delay when compared to the geometric means of the transfer functions.
This will be explored in depth within a follow-up work.

\subsection{BLR Contribution to Microlensing Signals}
\label{BLRMicrolensingSection}
The BLR is typically larger than the accretion disc, making it less sensitive and often used as a baseline in microlensing studies. 
However, the size scales of the AGN and the Einstein radius of microlenses are known to span multiple orders of magnitude~\citep[e.g.][]{Sluse11}.
If the BLR scales in size proportionally to the SMBH mass, some low mass AGN may have BLR with appropriate source sizes to be impacted by microlensing~\citep[e.g.][]{Hutsemekers19, Hutsemekers21}.

As described in Section~\ref{BLRsection}, \texttt{Amoeba} produces spatially resolved BLR based on their contribution to any described filter.
These projections are also flux distributions in the source plane and may be microlensed alongside the accretion disc. 
Furthermore, some parts of the BLR may reside in an over-density of caustics while other sections may be in an under-density.
The path of the BLR must be taken along the same path as the accretion disc in order to remain self-consistent.

\begin{figure}
    \centering
    \includegraphics[width=0.47\textwidth]{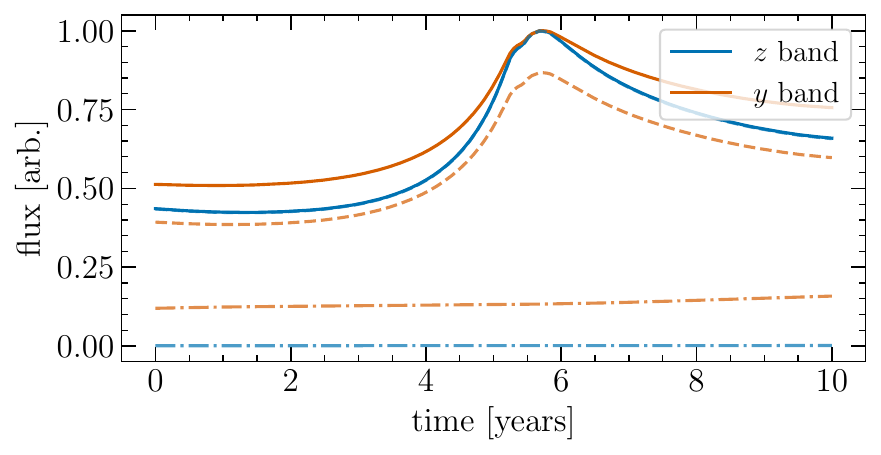}
    \caption{Microlensing light curve with BLR contributions as described in Section~\ref{SampleBLRContaminationSubsection}. The light curves are individually normalized, and the BLR is assumed to contribute 20 per cent of the accretion disc's flux in the $y$ band. The disc's contribution to each light curve is represented by the dashed line, while the BLR's contribution is the dashed-dotted line.}
    \label{FigdiscBLRMicrolensed}
\end{figure}

Fig.~\ref{FigdiscBLRMicrolensed} illustrates a microlensing light curve of the BLR set up in Section~\ref{SampleBLRContaminationSubsection} defined by parameters listed in Table~\ref{TableStreamlineParams} for an assumed $M_{\rm{BH}} = 10^{8} M_{\odot}$ alongside the accretion disc.
As illustrated in Fig.~\ref{ProjectedBLR}, the projected BLR takes on a ring-like appearance, making its light curve differ from the accretion disc's.
The contribution of the BLR to the total light curve depends on the emission line strength with respect to the underlying continuum.
In this case, we assume the emission line contributes 20 per cent of the accretion disc's flux in the $y$ band.
The $z$ band is scaled by the fractional emission with respect to the $y$ band's total emission, which is only 11.5 per cent in this example.
This leads to a negligible contribution in the $z$ band to this high magnification event.
The BLR's impact is highly model dependent so \texttt{Amoeba} aims to support each axi-symmetric geometry. 
These geometries may range from discs, to bi-conical flows, and even failed out flows.

We note that the accretion disc was modelled at effective wavelengths, while the BLR was modelled across the range of wavelengths corresponding to the width of optical filters.
Since the BLR's emission line is assumed to be sensitive to the line-of-sight velocity, the width of the filter becomes important to the regions that contribute to the emission.
On the other hand, we model the accretion disc such that it emits as a continuum.
Due to the broad emission of the continuum, the total radiation of the disc does not change rapidly between optical filters. 
The importance of integrating the disc's spectral flux density over an optical band is tested in Appendix~\ref{test_of_effective_wavelength_on_thin_disc}, where we determine the precise filter response leads to a negligible effect in the flux distribution for the black body continuum.

\subsection{Self-Consistent Light Curves}
\label{SelfConsistentSection}

\begin{figure*}
    \includegraphics[width=0.95\textwidth]{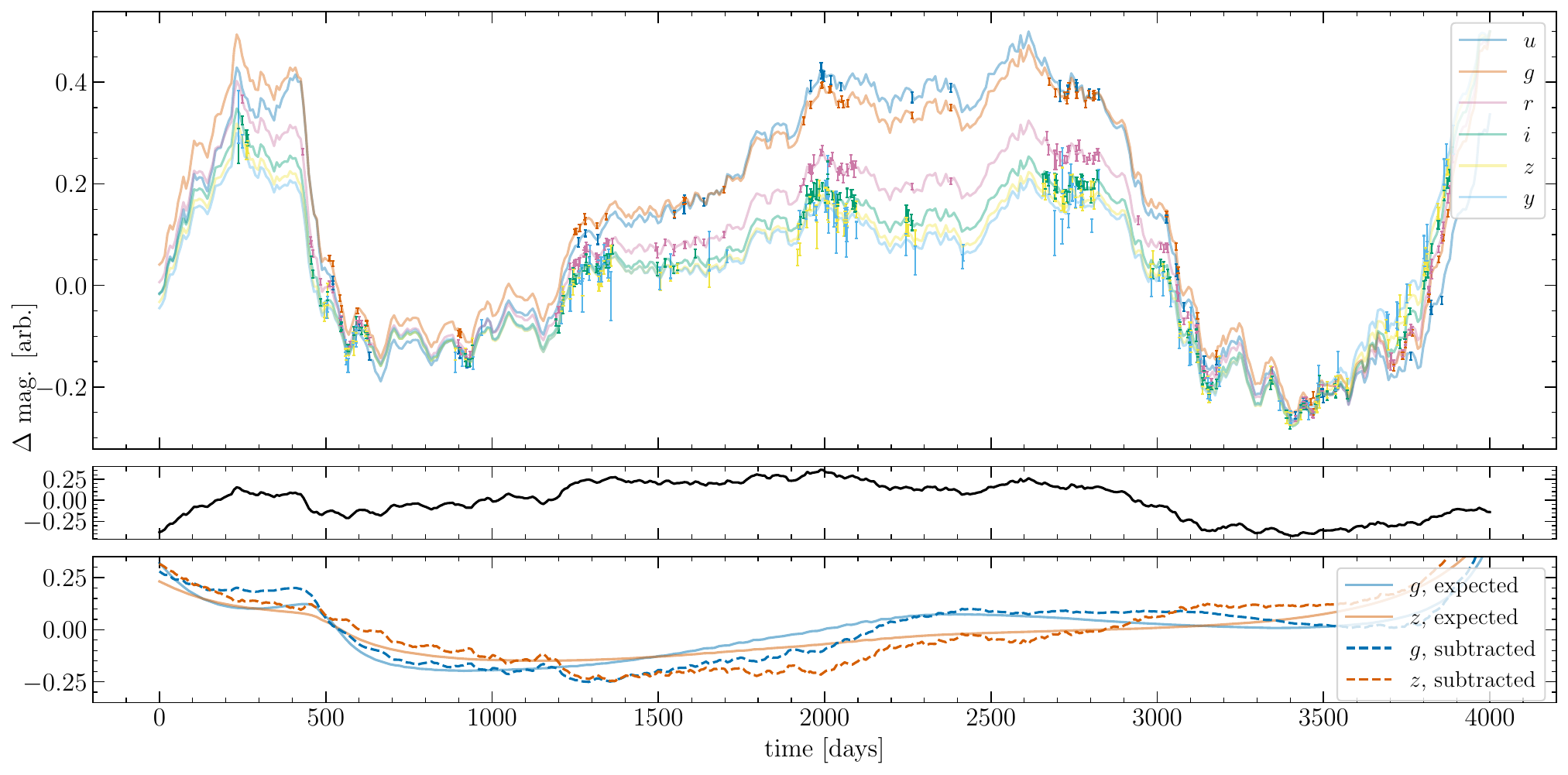}
    \caption{Self-consistent light curves incorporating the time variable accretion disc, BLR, and microlensing. The top panel illustrates a set of mock light curves simulated across LSST filters $u, g, r, i, z, y$ with mock observations. The second panel illustrates the intrinsic variability which was propagated through each component. The bottom panel illustrates the microlensing signals derived from two selected static accretion discs as well as the difference between the full light curve and the intrinsic variability. }
    \label{full_light_curve_decomposed}
\end{figure*}

To facilitate microlensing of time variable sources, \texttt{Amoeba} can construct snapshots of the source for each observation.
By propagating a signal through the source and microlensing each snapshot, higher order contributions to the variable signal may be simulated.
This leads to a deeper understanding of the models we use along with the interplay between them.

To create the surface flux distribution as observed at a certain time, the time-dependent temperature dictated by Equation (\ref{timedependenttemp}) is assumed to radiate in its rest frame following Equation (\ref{Planck}).
Due to the time lags described in Equation (\ref{timelageq}), the orientation of the accretion disc becomes important.
In the AGN's rest frame, the lamp-post model predicts temperature fluctuations as the driving signal evolves.
However, in the observer's frame, these rings appear distorted into isodelay paraboloids, which we define as the radiation pattern.

The radiation pattern is not an observable due to the fact that we cannot spatially resolve AGN accretion discs.
However, the radiation pattern can be important when microlensing caustic features are considered.
By generating the time variable surface flux distributions and microlensing these radiation patterns, we may relax the assumption that microlensing is stationary over a light curve.
This can be important for long light curves, where an AGN's intrinsic flux may vary upwards of a magnitude over time scales of years~\citep{Makarov24}.

With each of the component models that build up the AGN, we can produce a light curve representing each of their contributions.
In doing so, we generate a mock LSST light curve which consistently joins together the intrinsic driving signal, the reverberating accretion disc, the reverberating BLR, and microlensing in a realistic scenario.
The AGN is assumed to have an accretion disc, a BLR, and an intrinsic signal. 
Parameters of the accretion disc include $M_{\rm{BH}} = 10^{8} ~M_{\odot}, ~H_{*} = 10 ~r_{\rm{g}}, ~\eta_{\rm{x}} = 0.3, ~i = 25\degree, A_{*} = 0$.
The viscous profile of the accretion disc is taken to be the Novikov-Thorne profile with an Eddington ratio of 0.1.
The BLR parameters are taken as those in Table~\ref{TableStreamlineParams} and Section~\ref{SampleBLRContaminationSubsection} with a line emitting as a delta function at $\lambda_{\rm{rest}} = 486$ nm and the inclination changed to match this accretion disc.
As we are modelling a relatively face-on AGN, the torus is not assumed to contribute to the optical emission.
The intrinsic driving signal taken to have a power spectrum $\propto f^{-2.5}$.
Dense optical light curves are generated at daily cadence for each LSST filter ($u, g, r, i, z, y$) and used to drive the model bi-conical outflow from Section~\ref{BLRsection}.
The accretion disc continuum is calculated at each effective wavelength (e.g. see Appendix~\ref{test_of_effective_wavelength_on_thin_disc} for a comparison between effective wavelength and integration over the filter), while the range of wavelengths corresponding to more than 1 per cent response is used for the BLR emission.
The projected flux from the reverberating disc and BLR are convolved with a microlensing map at each observation timestamp, assuming $z_{\rm{s}} = 0.5, ~z_{\rm{l}} = 0.25, ~M_{\rm{ml}} = M_{\odot}, ~v_{\rm{eff}} = 800 \text{ km s}^{-1}$, to model the magnification due to microlensing.
The disc and BLR light curves are then combined assuming the emission line contributes 20 per cent of the accretion disc's flux in the $i$ filter.
A small portion of the BLR falls into the $r$ filter, while other filters are unaffected by this set of model parameters.
We do not apply a reddening model, and assume an average magnitude of 20.5 for each light curve.

We degrade and sample the dense light curves to simulate LSST-like observations.
We use LSST cadence and errors from \texttt{rubin\_sim}\footnote{\href{https://github.com/lsst/rubin\_sim}{https://github.com/lsst/rubin\_sim}} with the \texttt{baseline\_v2.1\_10yrs} rolling cadence for the Wide Fast Deep survey~\citep{LSST19}.
The light curves are then sampled following the procedure outlined in~\citet{Fagin24a} for each band.

Fig.~\ref{full_light_curve_decomposed} illustrates a representative set of mock light curves as observed in each LSST band for our model AGN.
The upper panel illustrates the dense light curves along with LSST sampling. 
Each light curve was taken to have an average magnitude of 20.5 in order to calculate observational uncertainties.
The second panel represents the stochastic driving signal which was propagated through the AGN.
The final panel illustrates the microlensing light curves for the $g$ and $z$ bands.
Solid lines represent the microlensing signal expected from a static source, while the dashed lines represent the microlensing signal derived from removing the intrinsic variability for selected bands from the dense light curve.
This is done to simulate the best case scenario where the intrinsic variability is removed from the total signal by the subtraction of two images.
The microlensing light curve labelled "subtracted" is very similar to the "expected" microlensing light curve, though there is deviation due to the intrinsic variability not being completely removed.

\section{Conclusions and Future Outlook}
\label{ConclusionSection}
We have presented \href{https://github.com/Henry-Best-01/Amoeba}{\texttt{Amoeba}}, a modular AGN modelling code which aims to combine intrinsic and extrinsic variability together across all components.
This initial release includes an accretion disc module with a temperature profile that is flexible and physically motivated, and converges smoothly to the thin disc profile.
Beyond this profile, any effective temperature mapping may be used with \texttt{Amoeba} allowing compatibility with other simulations.

A module for the BLR is included which is flexible enough to model any axi-symmetric region while retaining the local optimally emitting cloud model.
It is well suited to describe bi-conical flows which are accelerated along their poloidal direction as described by out flowing wind models.
The BLR module tracks velocity components to join Keplerian orbits with outflows.
By calculating line-of-sight velocities and assuming discrete emission lines, the contamination of broad filters may be determined.

A module for the diffuse continuum is included, which allows a user to model the increase in time lags between bands in addition to the traditional time lags associated with the transfer functions. 
This can be co-spatial with any other model components, and the average continuum responsivity spectrum may be computed using photo-ionization codes, allowing great flexibility for how the diffuse continuum is treated.
Currently, only a basic treatment of the diffuse continuum is implemented, but this will be the target of a future update as diffuse continuum models become more sophisticated.

A sub-module for torus obscuration is included, which naturally allows for differential obscuration.
This obscuration may be treated as proportional to a wavelength dependent factor and the column density.
The impact this has on the source (e.g. the accretion disc) is two-fold due to the natural wavelength dependent nature of AGN emission regions and the wavelength dependent attenuation.

The lamp-post model is built in to \texttt{Amoeba} and extended to allow off-axis sources to influence the accretion disc.
The corona is typically modelled as a point source above the SMBH and is allowed to drive variability in the accretion disc.
Any driving signal may be convolved with the transfer function in order to model the optical variability of the AGN.
Furthermore, a transfer function may be derived for any accretion disc temperature profile.
Off-axis events may be used to simulate flaring caused by exotic phenomena occurring in or near the accretion disc (e.g. a magnetic reconnection event~\citep{Dexter20}, a gamma ray burst due to a binary merger in the accretion disc~\citep{Mckernan12}, etc.).

The BLR is assumed to reverberate under the assumption that the accretion disc's continuum drives emission line variability.
This allows \texttt{Amoeba} to model the BLR's stochastic light curve along with its contamination to optical filters.
Mock observations may be simulated by combining the variable emission lines of the BLR to the continuum light curves.
We note that due to the BLR density and velocity distributions, the same emission line may have different apparent time lags in different filters.

\texttt{Amoeba} can perform microlensing simulations between magnification maps of appropriate size and the surface flux density of AGN components.
Each component of the AGN uses the SMBH as the point of reference in the convolution so they may be consistently microlensed.
Microlensing is known to be size dependent, so some components of the AGN will naturally be more sensitive to this phenomena than others.

\texttt{Amoeba} is designed to be modular with the intent of becoming an ever-growing AGN modelling tool.
Within the near future, some planned modules are:
\begin{enumerate}
    \item A complete torus model which includes spectral modelling of infrared emissions. 
    \item A method to incorporate realistic magneto-hydrodynamic simulations into the accretion disc model.
    \item A model for the AGN jet along with its ionization potential for the host galaxy.
\end{enumerate}
Furthermore, \texttt{Amoeba} is flexible and semi-analytic, giving it the power to derive flux distributions and reverberation properties from both analytic and non-analytic sources.
This is especially useful for studying non-homogeneous sources~\citep[e.g.][]{Zhou24}.

\texttt{Amoeba} has many applications, from building data sets for training machine learning algorithms to modelling multi-component self-consistent light curves.
In the context of high cadence caustic-crossing simulations, convolutional neural networks have been used to predict parameters of the AGN such as the mass of the SMBH, the inclination of the accretion disc, and its orientation with respect to the caustic fold~\citep{Best24}.
Such high cadence follow-up observations are expected to be triggered by the continuous monitoring by LSST and other wide-field surveys~\citep{Fagin24c}.
This trained network was then applied to archival data from QSO 2237+0305, the Einstein Cross, and predicted a SMBH mass within the previously measured values from literature.
In~\citet{Fagin24a}, \texttt{Amoeba} was used to simulate the optical response of the accretion disc due to X-ray variability in form of the DRW.
These continuous light curves were then used to produce LSST-like optical light curves, from which a generative machine learning method which solves underlying latent stochastic differential equations was used to simultaneously model the driving light curves and infer source parameters to higher effectiveness than traditional Gaussian process regression.
Furthermore, \texttt{Amoeba} is integrated into \href{https://github.com/LSST-strong-lensing/slsim}{\texttt{slsim}}, the strong lensing simulation pipeline for LSST.
\texttt{Amoeba} aims to be an open source, adaptable tool to join together the many models of AGN components into one coherent model.

\section*{Data Availability}

The data utilized in this paper are simulated and available upon reasonable request from the corresponding author. 
\texttt{Amoeba} is designed as an open source Python package and is available at \href{https://github.com/Henry-Best-01/Amoeba}{https://github.com/Henry-Best-01/Amoeba}.

\section*{Acknowledgements}
The authors thank the anonymous referee for comments which have greatly improved this manuscript and \texttt{Amoeba}.
This research was made possible by the generosity of Eric and Wendy Schmidt by recommendation of the Schmidt Futures program. 
H.B. acknowledges the GAČR Junior Star grant No. GM24-10599M for support.
The authors thank Giorgos Vernardos, Timo Anguita, and Dominique Sluse for helpful discussion regarding gravitational lensing, Saavik Ford and Barry McKernan for helpful discussion regarding the BLR as well as the prospect of non-uniform accretion discs.
H.B. thanks Simon Birrer and Narayan Khadka for helpful discussion regarding the code, and Michal Zaja\v{c}ek for helpful discussion regarding variability.
The following Python packages and modules were used in this work: \texttt{Numpy}\footnote{\url{https://numpy.org}}, 
\texttt{Scipy}\footnote{\url{https://scipy.org}}, 
\texttt{Astropy}\footnote{\url{https://www.astropy.org}}, 
\texttt{Scikit-Image}\footnote{\url{https://scikit-image.org}},
\texttt{Matplotlib}\footnote{\url{https://matplotlib.org}}, 
\texttt{Sim5}\footnote{\url{https://github.com/mbursa/sim5/}}, 
\texttt{Speclite}\footnote{\url{https://speclite.readthedocs.io/en/latest/overview.html}}

\bibliographystyle{mnras}
\bibliography{bib}

\appendix

\section{Accretion Disc Effective Temperature Profile}
\renewcommand{\thefigure}{A.\arabic{figure}}
\renewcommand{\thetable}{A.\arabic{table}}
\setcounter{figure}{0}  
\setcounter{table}{0} 

\label{appendix_temp_profile}

The temperature profile of the accretion disc is intricately connected to its emission and variability properties.
In this appendix we join together modifications to the thin disc temperature profile which allow for greater flexibility in models.
This section assumes constant values of $M_{\rm{BH}} = 10^{8} M_{\odot}, r_{\rm{in}} = r_{\rm{ISCO}} \equiv 6 r_{\rm{g}}$, and an Eddington ratio of 0.15.

The thin disc considers the effect of black-body radiation due to viscous accreted material.
The lamp-post heated accretion disc model goes beyond this to include extra thermal energy from the lamp-post~\citep{Sergeev05, Cackett07}.
This is assumed to originate from the corona, where highly energetic photons are emitted at X-ray wavelengths and are allowed to irradiate the disc.
The modified temperature profile due to this mean irradiating flux is defined as in~\citet{Cackett07}:
\begin{equation}
    \label{lamppostdisc}
    T^{4}_{\rm{D+L}}(R) = \left[ T^{4}_{\rm{visc}}(R) + \left( \frac{(1-A_{*}) L_{\rm{x}}}{4\pi \sigma_{\rm{sb}} r^{2}_{*}}\right) \cos(\theta_{\rm{x}}) \right] 
\end{equation}
where $A_{*}$ is the albedo of the accretion disc, $L_{\rm{x}} \equiv \eta_{\rm{x}} \dot{M} c^{2}$ is the lamp-post flux with efficiency parameter $\eta_{\rm{x}}$, $r_{*}$ is the distance from the lamp-post, and $\cos(\theta_{\rm{x}})$ is the angle of incidence of the illuminating flux on the disc.
For a flat accretion disc, $r_{*} \equiv \sqrt{H_{*}^{2} + r_{*}^{2}}$ and $\cos(\theta_{\rm{x}}) = H_{*} / r_{*}$.
Some authors include thermal heating from the second lamp-post on the opposite side of the accretion disc in this equation, but we note that the energy transfer through the disc (especially discs that aren't razor thin) may not be instantaneous.
Some work has been done on more advanced accretion discs, such as incorporating the reflectivity of the disc~\citep[e.g.][]{Kammoun19} or rimmed and rippled accretion disc geometries~\citep[e.g.][]{Starkey23}.
Within this work, we assume flat geometries with constant albedo for simplicity and note that \texttt{Amoeba} has the capability to model more complex accretion discs.

\begin{figure}
    \centering
    \includegraphics[width=0.47\textwidth]{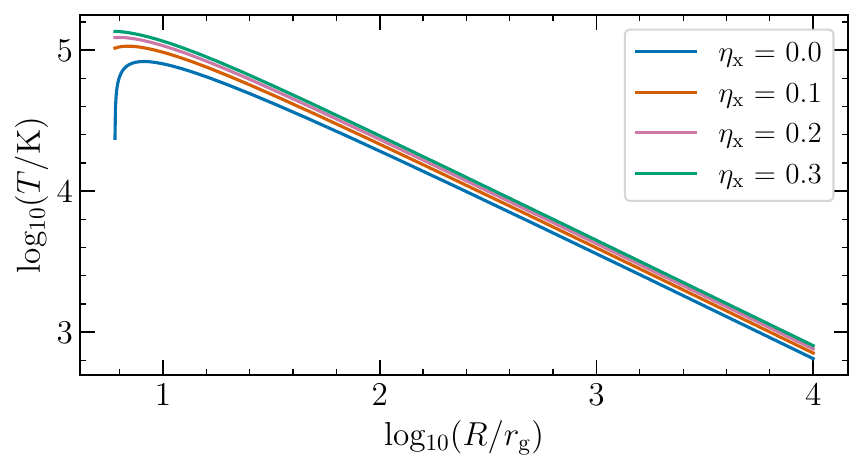}
    \caption{Temperature profiles for the lamp-post heated accretion disc model. The height of the irradiating source is chosen to be 6 $r_{\rm{g}}$ above the central axis in all cases. $\eta_{\rm{x}}$ parameterizes the strength of corona as a fraction of the total energy due to accretion.}
    \label{LampPostProfiles}
\end{figure}

Fig.~\ref{LampPostProfiles} illustrates the effect of increasing $\eta_{\rm{x}}$ on the temperature profile of this model. 
In each case, the source is placed on the axis of symmetry at a height of $6 r_{\rm{g}}$.
The temperature at all radii is elevated without an effect on the asymptotic power law of $T \propto R^{-3/4}$.

Most accretion disc profiles assume a constant accretion rate throughout the disc.
There is also the possibility that some accreted matter feeds into winds in the form of an out flow.
This has been explored and modelled in various ways~\citep[e.g.][and references therein]{You16}.
For the case of a viscous accretion disc with wind, the temperature profile may have an effective change in the power law dependence.
We describe the modified temperature profile due to this affect as~\citep{Sun19}:
\begin{equation}
  \label{discWind}
  T^{4}_{\rm{D+W}}(R) = \left(\frac{3GM_{\rm{BH}} \dot{M}(r_{\rm{in}}) \left(1 - \sqrt{r_{\rm{in}} / R}\right)}{8\pi \sigma_{\rm{sb}} (r_{\rm{in}})^{3}} \right) \left(\frac{R}{r_{\rm{in}}}\right)^{(\beta_{\rm{w}} - 3)}
\end{equation}
where $\beta_{\rm{w}}$ defines the strength of the wind and $\dot{M}(r_{\rm{in}})$ is the accretion rate at the ISCO.
This form arises under the assumption the accretion rate varies throughout the disc as $\dot{M}(R) = \dot{M}(r_{\rm{in}}) (R/r_{\rm{in}})^{\beta_{\rm{w}}}$, and does not include irradiation by the corona.
The wind removes accreted material as it travels inward toward $r_{\rm{in}}$, where the wind vanishes. 

\begin{figure}
    \centering
    \includegraphics[width=0.47\textwidth]{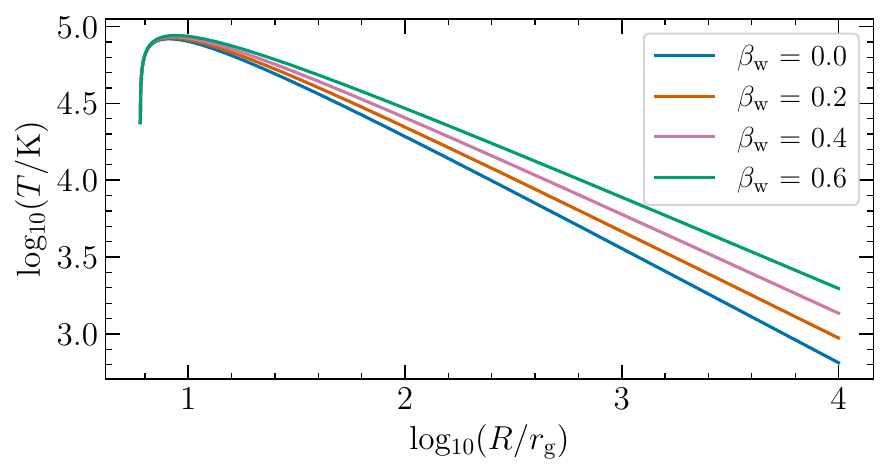}
    \caption{Temperature profile assuming the thin disc with wind profile following Equation (\ref{discWind}) at various $\beta_{\rm{w}}$. The $\beta_{\rm{w}} = 0$ case corresponds to the thin disc solution. $\beta_{\rm{w}}$ acts to adjust the asymptotic slope of the temperature profile by giving the accretion rate a radial dependence.}
    \label{ProfileWithWinds}
\end{figure}

Various mechanisms may produce this dependence of the accretion rate at different radii~\citep[See][for a brief review]{Proga07}.
We do not explore the particular mechanisms or the extent of the wind region, and instead opt to adopt this functional form into the temperature profile.
Fig.~\ref{ProfileWithWinds} illustrates the temperature profiles for varying levels of $\beta_{\rm{w}}$, and highlights the smooth connection to the thin disc temperature profile for $\beta_{\rm{w}}$ = 0.
The maximum temperature of the disc is not significantly affected as this model primarily adjusts larger scales. 
Its asymptotic power law dependence becomes $(\beta_{\rm{w}} - 3) / 4$, allowing for shallower accretion disc temperature profiles.
It should be noted that this changes the total power of the accretion disc, causing the Eddington ratio to no longer be conserved when $\beta_{\rm{w}}$ is adjusted.

\begin{figure}
    \centering
    \includegraphics[width=0.47\textwidth]{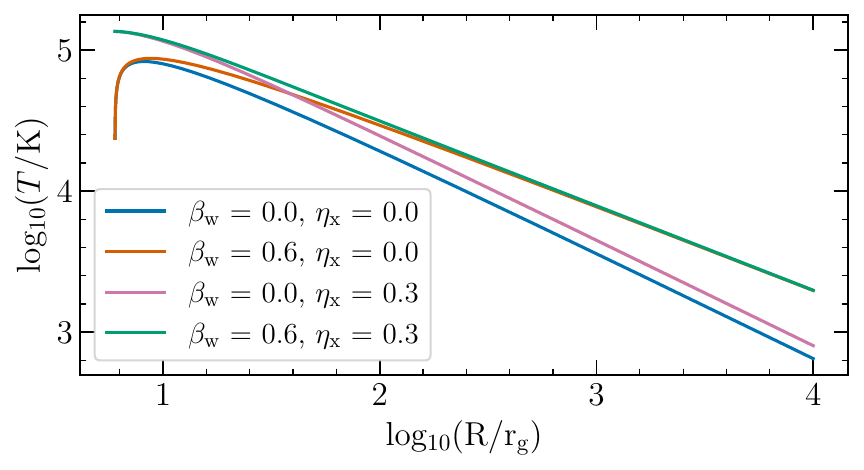}
    \caption{Various accretion disc temperature profiles as a function of radius. The $\beta_{\rm{w}}$ = 0, $\eta_{\rm{x}}$ = 0 case corresponds to the thin disc solution. Large values of $\beta_{\rm{w}}$ and $\eta_{\rm{x}}$ are used to accentuate their impact on the temperature profiles.}
    \label{ProfileComparison}
\end{figure}

The temperature profiles defining Equations (\ref{discWind}) and (\ref{lamppostdisc}) represent different physical processes.
If we make the assumption that these processes may be joined together, we will have the resulting temperature profile:
\begin{equation}
\begin{split}
  \label{fulltempprofile}
  T^{4}(R) = \left(\frac{3GM_{\rm{BH}}\dot{M}(r_{\rm{in}}) \left(1 - \sqrt{r_{\rm{in}} /R}\right)}{8\pi \sigma_{\rm{sb}} (r_{\rm{in}})^{3}} \right) &\left(\frac{R}{r_{\rm{in}}}\right)^{(\beta_{\rm{w}}-3)} \\
  &+ \left( \frac{(1-A_{*}) L_{\rm{x}}}{4\pi \sigma_{\rm{sb}} r^{2}_{*}}\right) \cos(\theta_{\rm{x}})
\end{split}
\end{equation}
This profile converges to the thin disc profile of Equation (\ref{Thindisc}) when $\beta_{\rm{w}} = 0$ and $L_{\rm{x}} = 0$. 
Holding $\beta_{\rm{w}} = 0$ and increasing $L_{\rm{x}}$ allows us to model the lamp-post irradiated disc, following Equation (\ref{LampPostProfiles}).
Adjusting $\beta_{\rm{w}}$ and setting $L_{\rm{x}} = 0$ brings us to the thin disc with wind profile defined by Equation (\ref{discWind}).
This is a more flexible temperature profile than the thin disc profile, and still remains analytic and physically motivated.

Fig.~\ref{ProfileComparison} illustrates the smooth connection between these profiles for each parameter combination of $\beta_{\rm{w}} = \{0, 0.6\}$ and $\eta_{\rm{x}} = \{0, 0.3\}$.
It is apparent that the thin disc sets the lower temperature limit in all cases.
The lamp-post strength adjusts the overall temperature and the wind parameter determines the asymptotic slope.
We note that the relativistic Novikov-Thorne viscous profile is also incorporated in \texttt{Amoeba}.

\section{BLR Parameterization}
\renewcommand{\thefigure}{B.\arabic{figure}}
\renewcommand{\thetable}{B.\arabic{table}}
\setcounter{figure}{0}  
\setcounter{table}{0} 

The BLR is modelled as a bi-conical out flow bounded by two streamlines within this work. 
Parameters of each streamline are defined in Table~\ref{TableStreamlineParams}, which correspond to the accelerated outflow defined in Equation (\ref{PoloidalVelocityEqu}) with $\alpha_{\rm{w}} = 1$.
The maximum height of the BLR, $H_{\rm{max}}$ is the same for both streamlines.
These parameters are chosen to accentuate \texttt{Amoeba}'s ability to model the high Keplerian velocity (e.g. $R_{\rm{l, SL1}} = 100 r_{\rm{g}}$) with comparable out flowing velocity (e.g. $v_{\infty, \rm{SL1}} = 0.2 c$). 

\begin{table}
    \centering
    \caption{Parameters for the example BLR used throughout this work.}
    \begin{tabular}{l|c|c}
         Parameter & Value & Units\\
    \hline\hline
         $R_{\rm{l, SL1}}$ & 100 & $r_{\rm{g}}$\\
         $\theta_{\rm{SL1}}$ & 10 & $\degree$\\
         $v_{0, \rm{SL1}} $  & 0 & $c$\\
         $R_{v, \rm{SL1}}$ & 200 & $r_{\rm{g}}$ \\
         $v_{\infty, \rm{SL1}} $ & 0.2 & $c$\\
         $R_{\rm{l, SL2}}$ & 400 & $r_{\rm{g}}$\\
         $\theta_{\rm{SL2}}$ & 30 & $\degree$\\
         $v_{0, \rm{SL2}} $ & 0 & $c$\\
         $R_{v, \rm{SL2}}$ & 500 & $r_{\rm{g}}$ \\
         $v_{\infty, \rm{SL2}} $ & 0.15 & $c$\\
         $H_{\rm{max}}$ & 1000 & $r_{\rm{g}}$\\
    \end{tabular}
    \label{TableStreamlineParams}
\end{table}

\section{Effective Wavelength Verses Total Filter Response}
\renewcommand{\thefigure}{C.\arabic{figure}}
\renewcommand{\thetable}{C.\arabic{table}}
\setcounter{figure}{0}  
\setcounter{table}{0} 
\label{test_of_effective_wavelength_on_thin_disc}

\begin{figure}
    \centering
    \includegraphics[width=0.47\textwidth]{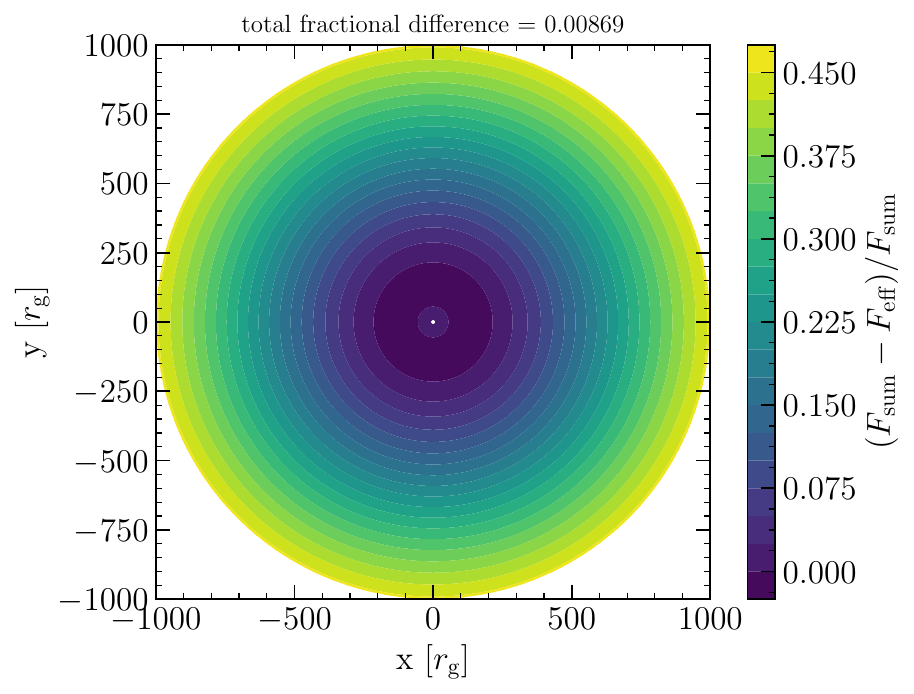}
    \caption{Fractional deviation between the effective flux distribution and summed flux distribution of a thin accretion disc with $M_{\rm{BH}} = 10^{8} M_{\odot}$ at $z_{\rm{s}}$ = 2.}
    \label{accretion_disk_residual_map}
\end{figure}

We conduct a short investigation on the significance of calculating the accretion disc's flux using the effective wavelength compared to using an expected filter response curve within the context of \texttt{Amoeba}.
We use the \href{https://speclite.readthedocs.io/en/latest/overview.html}{\texttt{Speclite}} package~\citep{speclite23} to model filter responses.
From the \texttt{Speclite} package, we load the filter associated with the current LSST $i$-band throughput, labelled "lsst2023-i".
This filter gives the associated response for each angstrom spanning the optical wavelengths.
The total wavelength range for responses greater than 1 per cent is 6725 to 8358 angstroms, with an effective wavelength of 7579 angstroms and an average response of 0.476.

The accretion disc is assumed to be a thin disc viewed face-on at $i = 0\degree$, accreting at Eddington ratio 0.1 with $r_{\rm{in}} = 6 r_{\rm{g}}$ and $r_{\rm{out}} = 1000 r_{\rm{g}}$.
We test the difference in calculated flux from the accretion disc with $M_{\rm{BH}} = 10^{8} M_{\odot}$, located at $z_{\rm{s}} = 2$. 

We note that the effective flux distribution $F_{\rm{eff}}$ and the summed flux distribution $F_{\rm{sum}}$ of the accretion disc must be calculated in a comparable manner.
For the effective flux distribution, the spectral radiance is calculated at the effective wavelength, weighted by the average throughput, and multiplied by the wavelength range of the filter.
For the summed flux distribution, the spectral radiance is calculated at each discrete wavelength, weighted by the throughput at that wavelength and the spectral resolution of 1 angstrom, then summed over.
The summed flux distribution represents the numerical integral that should be considered when modelling fluxes from the accretion disc's spectral radiance.

\begin{figure}
    \centering
    \includegraphics[width=0.47\textwidth]{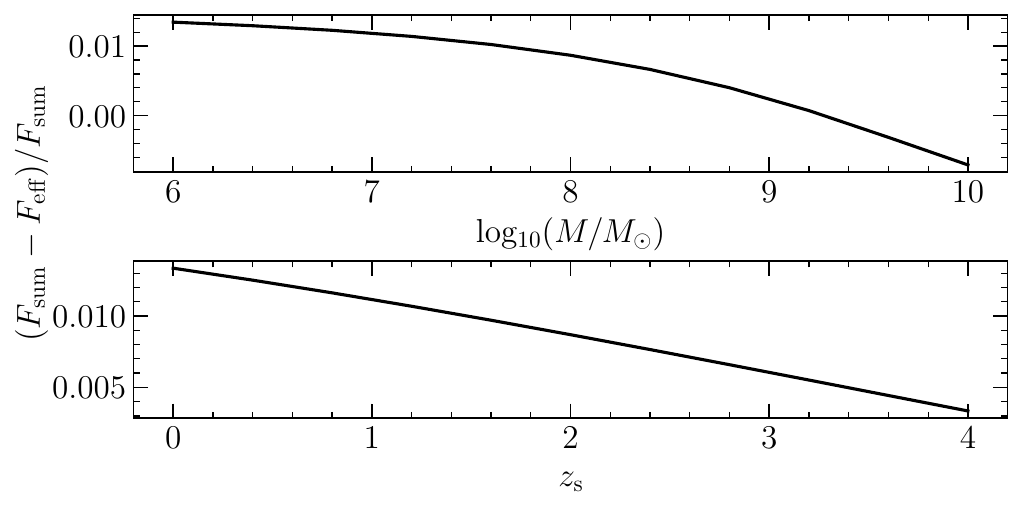}
    \caption{Relative change in integrated flux of the accretion disc when calculated at the effective wavelength and when integrated over a \texttt{speclite} filter for the LSST $i$ band.}
    \label{fig_flux_deviation_for_continuous_vs_discrete}
\end{figure}

Fig.~\ref{accretion_disk_residual_map} illustrates the distribution of fractional residuals defined as $(F_{\rm{sum}} - F_{\rm{eff}}) / F_{\rm{sum}}$.
The greatest deviations between $F_{\rm{sum}}$ and $F_{\rm{eff}}$ occur at large radii where the radiation density is low.
Fig.~\ref{fig_flux_deviation_for_continuous_vs_discrete} illustrates the deviation in total flux by integrating each distribution of $F_{\rm{sum}}$ and $F_{\rm{eff}}$ for multiple values of $M_{\rm{BH}}$ and $z_{\rm{s}}$.
For typical accretion discs, the deviation between calculating the flux at the effective wavelength compared to integrating over the filter is less than 1 per cent. 
For low mass, low redshift sources, the deviation between calculations can approach $\sim$ 2 per cent.
The top panel represents how this deviation evolves with mass for fixed $z_{\rm{s}} = 2$, while the bottom panel represents the deviation's evolution with redshift for fixed $M_{\rm{BH}} = 10^{8} M_{\odot}$.
We do not find a significant deviation, and conclude that calculating the accretion disc's flux at the filter's effective wavelength is an efficient and accurate estimation of the integrated flux over an optical band for black body radiation.

We also test this effect in the calculation of accretion disc transfer functions.
The model is identical to the previous: we load the filter associated with the current LSST $i$-band response labelled "lsst2023-i" and use its discrete form.
The transfer functions are then calculated for $M_{\rm{BH}} = 10^{8} M_{\odot}$ at $z_{\rm{s}} = 2$ for the effective wavelength and the range of wavelengths.
Normalization occurs only after the transfer functions are combined and weighted by the response.

\begin{figure}
    \centering
    \includegraphics[width=0.47\textwidth]{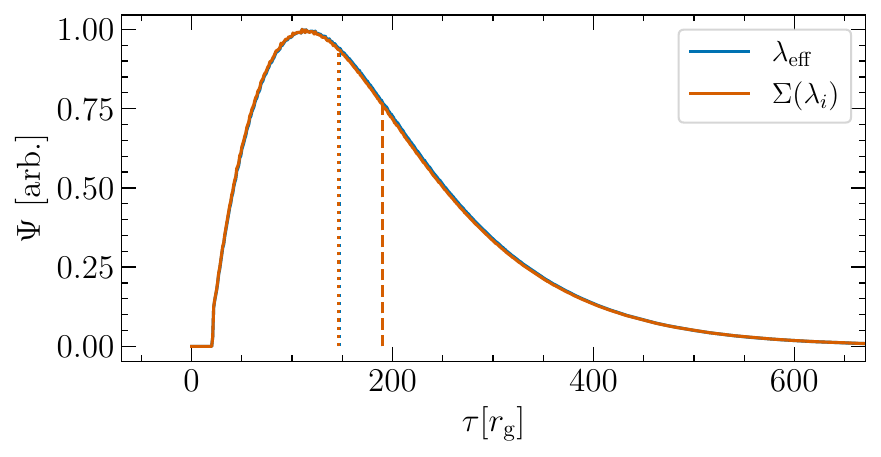}
    \caption{Transfer function calculated for an accretion disc at the effective wavelength and over the \texttt{speclite} "lsst2023-i" filter.}
    \label{discrete_vs_continuous_tfs_for_disk}
\end{figure}

Fig.~\ref{discrete_vs_continuous_tfs_for_disk} illustrates the outcome of this test for the expected response of the LSST $i$ band.
We find that the difference between the computed transfer functions is negligible for both the geometric mean and centroid time delays.
We conclude that calculating the accretion disc's transfer function at the filter's effective wavelength is efficient and accurate for optical wavelengths.

\bsp	
\label{lastpage}

\end{document}